\documentclass[12pt]{article}

\usepackage{graphicx}

\def\ltwid{\mathrel{\raise.3ex\hbox{$<$\kern-.75em\lower1ex\hbox{$\sim$}}}}
\def\gtwid{\mathrel{\raise.3ex\hbox{$>$\kern-.75em\lower1ex\hbox{$\sim$}}}}

\def\Fint{\rlap{$\Biggl\rfloor$}\Biggl\lceil}
\def\square{\kern1pt\vbox{\hrule height 1.2pt\hbox{\vrule width 1.2pt\hskip 3pt
   \vbox{\vskip 6pt}\hskip 3pt\vrule width 0.6pt}\hrule height 0.6pt}\kern1pt}
\def\overleftrightarrow#1{\vbox{\ialign{##\crcr
     $\leftrightarrow$\crcr\noalign{\kern-1pt\nointerlineskip}
     $\hfil\displaystyle{#1}\hfil$\crcr}}}

\begin{document}

\begin{titlepage}

\begin{flushright}
CCTP-2011-27 \\ UFIFT-QG-11-07
\end{flushright}

\vskip 1cm

\begin{center}
{\bf Generalizing the ADM Computation to Quantum Field Theory}
\end{center}

\begin{center}
P. J. Mora$^{1*}$, N. C. Tsamis$^{2\dagger}$ and R. P.
Woodard$^{1,\ddagger}$
\end{center}

\begin{center}
\it{$^{1}$ Department of Physics, University of Florida \\
Gainesville, FL 32611, UNITED STATES}
\end{center}

\begin{center}
\it{$^{2}$ Institute of Theoretical Physics \& Computational Physics \\
Department of Physics University of Crete \\
GR-710 03 Heraklion, HELLAS}
\end{center}

\begin{center}
ABSTRACT
\end{center}
The absence of recognizable, low energy quantum gravitational
effects requires that {\it some} asymptotic series expansion be
wonderfully accurate, but the correct expansion might involve
logarithms or fractional powers of Newton's constant. That would
explain why conventional perturbation theory shows uncontrollable
ultraviolet divergences. We explore this possibility in the context
of the mass of a charged, gravitating scalar. The classical limit of
this system was solved exactly in 1960 by Arnowitt, Deser and
Misner, and their solution does exhibit nonanalytic dependence on
Newton's constant. We derive an exact functional integral
representation for the mass of the quantum field theoretic system,
and then develop an alternate expansion for it based on a correct
implementation of the method of stationary phase. The new expansion
entails adding an infinite class of new diagrams to each order and
subtracting them from higher orders. The zeroth order term of the
new expansion has the physical interpretation of a first quantized
Klein-Gordon scalar which forms a bound state in the gravitational
and electromagnetic potentials sourced by its own probability
current. We show that such bound states exist and we obtain
numerical results for their masses.

\begin{flushleft}
\end{flushleft}

\begin{flushleft}
$^*$ e-mail: pmora@phys.ufl.edu \\
$^{\dagger}$ e-mail: tsamis@physics.uoc.gr \\
$^{\ddagger}$ e-mail: woodard@phys.ufl.edu
\end{flushleft}

\end{titlepage}

\section{Introduction}\label{intro}

The problem of quantum gravity is that perturbative loop corrections
to quantum general relativity contain ultraviolet divergences that
can only be absorbed by adding higher derivative counterterms which
would make the universe decay instantly \cite{RPW1}. The divergence
problems are well known and ubiquitous:
\begin{itemize}
\item{Einstein + scalar requires a bad counterterm at one
loop order \cite{HV};}
\item{The same unacceptable one loop counterterm is also needed for
Einstein + Maxwell \cite{DN1}, Einstein + Dirac \cite{DN2} and
Einstein + Yang-Mills \cite{ DTN};}
\item{The Einstein theory by itself requires an unacceptable
counterterm at two loop order \cite{GS}; and}
\item{Although supergravity is on-shell finite at two loop order
\cite{DKS}, and explicit computation shows that $N=8$ supergravity
is on-shell finite at three \cite{Bern1} and four loop order
\cite{Bern2}, all supergravity models are expected to require
unacceptable counterterms by seven loop order \cite{Stelle}.}
\end{itemize}

Quantum gravity can of course be used as an effective field theory
by treating the bad counterterms as perturbations and then
restricting to low energy predictions \cite{Donoghue} which are
insensitive to them. If the Asymptotic Safety scenario
\cite{Weinberg} is realized, it might even be that the escalating
series of perturbative counterterms does not spoil predictivity at
energies below the Planck scale. However, neither approach provides
a fundamental resolution.

The problem arises from the tension between four facts \cite{RPW1}:
\begin{itemize}
\item{{\it Continuum Field Theories} possess an infinite number of modes;}
\item{{\it Quantum Mechanics} requires each mode to have a minimum amount of
energy;}
\item{{\it General Relativity} stipulates that stress-energy is the source
of gravitation; and}
\item{{\it Perturbation Theory} simply adds up the contribution from each
mode at lowest order.}
\end{itemize}
One or more of these principles must be sacrificed, and a little
thought suggests focussing on the last two. There does not seem any
way of disputing the experimental confirmation of quantum mechanics
in the matter sector which is responsible for the lowest order
divergences of quantum general relativity. And inflationary
cosmology makes nonsense of any attempt to invoke a nonzero physical
cutoff length. Inflation predicts that the universe has expanded by
the staggering factor of at least $10^{52}$ \cite{Linde}, so if the
physical cutoff is at the Planck length today then it must have been
about $10^{-90}~{\rm m}$ during primordial inflation. But fossilized
quantum gravitational effects from primordial inflation have been
measured with a fractional strength of about $10^{-10}$ \cite{WMAP},
which is inconsistent with so small a physical cutoff length.

Superstring theory can be viewed as an attempt to preserve the
validity of perturbation theory by sacrificing general relativity.
We wish here to investigate the alternative: that the problems of
quantum general relativity derive from using conventional
perturbation theory. We disavow any intention of seeking the exact
solution. There is so far no example of an interacting quantum field
theory in $D = 3 + 1$ dimensions which can be solved exactly, and
all experience with classical field theory suggests that general
relativity is an unlikely candidate to be the first one. What
interests us instead is the possibility that quantum general
relativity has a perfectly finite asymptotic series expansion which
is simply not given by conventional perturbation theory.

The conventional result for the expectation value of a quantum
gravity observable $\mathcal{O}$ with characteristic length $R$ is
assumed to take the form,
\begin{equation}
\Bigl\langle \Omega \Bigl\vert \mathcal{O} \Bigr\vert \Omega
\Bigr\rangle = \Bigl( {\rm Tree\ Order}\Bigr) \Biggl\{ 1 +
\sum_{\ell=1}^{\infty} a_{\ell} \Bigl( \frac{\hbar G}{c^3
R^2}\Bigr)^{\ell} \Biggr\} \; , \label{usual}
\end{equation}
where $G$ is Newton's constant, $\hbar$ is Planck's constant and $c$
is the speed of light. Support for this form is adduced from the
fact that quantum gravity has no observable effects at low energies.
Even for the smallest distances ever probed, $R \sim 10^{-19}~{\rm
m}$, the loop counting parameter is minuscule, $\hbar G/(c^3 R^2)
\sim 10^{-32}$. But the same thing would be true of a series that
incorporates fractional powers or logarithms such as,
\begin{equation}
\Bigl\langle \Omega \Bigl\vert \mathcal{O} \Bigr\vert \Omega
\Bigr\rangle = \Bigl( {\rm Tree\ Order}\Bigr) \Biggl\{ 1 +
\sum_{\ell=1}^{\infty} \sum_{m=0}^{\ell} a_{\ell m} \Bigl(
\frac{\hbar G}{c^3 R^2}\Bigr)^{\ell} \Bigl[\ln\Bigl( \frac{\hbar
G}{c^3 R^2}\Bigr)\Bigr]^m \Biggr\} \; . \label{possible}
\end{equation}
If the actual asymptotic expansion of quantum gravity were to take
the form (\ref{possible}) then loop effects at $R \sim 10^{-19}~{\rm
m}$ would still be suppressed by unobservably small powers of the
parameter,
\begin{equation}
\Big(\frac{\hbar G}{c^3 R^2}\Bigr) \ln\Bigl(\frac{\hbar G}{c^3 R^2}
\Bigr) \Biggl\vert_{R \sim 10^{-19}~{\rm m}} \sim -10^{-30} \; .
\end{equation}
However, trying to force the putative series (\ref{possible}) into
the assumed form (\ref{usual}) would result in logarithmically
divergent coefficients $a_{\ell}$, which is exactly what explicit
computations reveal.

The incorporation of such nonanalytic terms into an asymptotic
expansion occurs even for very simple physical systems. Consider the
canonical partition function for a non-interacting, highly
relativistic particle of mass $m$ in a three dimensional volume $V$
at temperature $k_B T = 1/\beta$,
\begin{eqnarray}
Z & = & {V \over 2 \pi^2 \hbar^3} \int_0^{\infty} \!\! dp \, p^2
\exp\Bigl[- \beta \sqrt{p^2 c^2 \!+\! m^2 c^4} + \beta m c^2\Bigr]
\; , \\
& = & {V \over 2 \pi^2 \hbar^3 c^3} \int_0^{\infty} \!\! dK \, (K
\!+\! m c^2) \; \sqrt{K^2 \!+\! 2 K m c^2} \, \exp(-\beta K) \; .
\qquad
\end{eqnarray}
When the rest mass energy is small compared to the thermal energy it
ought to make sense to expand in the small parameter $x \equiv \beta
m c^2$. But straightforward perturbation theory fails,
\begin{eqnarray}
\lefteqn{Z = {V \over 2 \pi^2} \Bigl({k_B T \over \hbar c}\Bigr)^3
\int_0^{\infty} \!\!\!\! dt \, t^2 e^{-t} \Bigl(1 \!+\!
\frac{x}{t}\Bigr) \sqrt{1 \!+\! 2 \frac{x}{t}} \; , } \\
& & \hspace{-.5cm} = {V \over 2 \pi^2} \Bigl({k_B T \over \hbar
c}\Bigr)^3 \!\!\! \int_0^{\infty} \!\!\!\! dt \, t^2 e^{-t}
\Biggl\{1 \!+\! 2 \frac{x}{t} \!+\! \frac12 \Bigl(\frac{x
}{t}\Bigr)^2 \!\!\!-\! \sum_{n=3}^{\infty} {(n \!-\! 3)(2n \!-\!
5)!! \over n!} \Bigl(-\frac{x}{t}\Bigr)^n\Biggr\} . \qquad
\label{wrong}
\end{eqnarray}
From expression (\ref{wrong}) it seems as though the term of order
$x^3$ vanishes, and that the higher terms have increasingly
divergent coefficients with oscillating signs. In fact the $x^3$
term is non-zero, and the apparent divergences merely signal
contamination with logarithms,
\begin{equation}
Z  =  {V \over 2 \pi^2} \Bigl({k_B T \over \hbar c}\Bigr)^3
\Biggl\{2 \!+\! 2 x \!+\! \frac12 x^2 \!-\! \frac16 x^3 \!-\!
\frac1{48} x^4 \ln(x) \!+\! O\Bigl(x^4\Bigr)\Biggr\} \; .
\end{equation}
Just as we suspect is the case for quantum gravity, the partition
function has an excellent expansion for small $x = \beta m c^2$; the
terms after order $x^2$ are indeed smaller than $x^2$, they just are
not as small as one naively thinks.

Rather than attempting to develop a new expansion for an arbitrary
quantum gravity observable, we restrict attention here to the
self-energy of a charged, gravitating particle. An exact result for
the $\hbar = 0$ limit of this system was obtained in 1960 by
Arnowitt, Deser and Misner (ADM) \cite{ADM1}. Their work provides
strong support both for the possibility that negative gravitational
interaction energy cancels divergences, and for the possibility that
the correct asymptotic expansion involves nonanalytic dependence on
Newton's constant. We review this evidence in section 2. In section
3 we propose an alternate expansion scheme for the self-energy of a
quantum field-theoretic particle. How the new expansion reshuffles
the diagrams of conventional perturbation theory is worked out in
section 4. We discuss the 0th order term of the new expansion in
section 5. Our conclusions comprise section 6.

\section{The ADM Computation}

Arnowitt, Deser and Misner showed that perturbation theory breaks
down in computing the self-energy of a classical, charged,
gravitating point particle \cite{ADM1}. It is simplest to model the
particle as a stationary spherical shell of radius $R$, charge $e$
and bare mass $M_0$. In Newtonian gravity its rest mass energy would
be,
\begin{equation}
M_R c^2 = M_0 c^2 + {e^2 \over 8 \pi \epsilon_0 R} - {G M_0^2 \over
2 R} \; .
\end{equation}
It turns out that all the effects of general relativity are
accounted for by assuming it is the full mass $M_R$ which
gravitates, rather than $M_0$ \cite{ADM1},
\begin{equation}
M_{R} = M_0 + {e^2 \over 8 \pi \epsilon_0 R c^2} - {G M^2_{R} \over
2 R c^2} = {R c^2 \over G} \sqrt{1 + {2 G \over R c^2} \Bigl(M_0 +
{e^2 \over 8 \pi \epsilon_0 R c^2}\Bigr)} - \frac{R c^2}{G} \; .
\end{equation}

The perturbative result is obtained by expanding the square root,
\begin{equation}
M_{\rm pert} = M_0 + {e^2 \over 8 \pi \epsilon_0 R c^2} +
\sum_{n=2}^{\infty} {(2n-3)!! \over n!} \; \Biggl(-{G \over R
c^2}\Biggr)^{n-1} \Biggl(M_0 + {e^2 \over 8 \pi \epsilon_0 R
c^2}\Biggr)^n \; , \label{pert}
\end{equation}
and shows the oscillating series of increasingly singular terms
characteristic of the previous examples. The alternating sign
derives from the fact that gravity is attractive. The positive
divergence of order $e^2/R$ evokes a negative divergence or order $G
e^4/R^3$, which results in a positive divergence of order $G^2
e^6/R^5$, and so on. The reason these terms are increasingly
singular is that the gravitational response to an effect at one
order is delayed to a higher order in perturbation theory.

The correct result is obtained by taking $R$ to zero before
expanding in the coupling constants $e^2$ and $G$,
\begin{equation}
\lim_{R \rightarrow 0} M_{R} = \Biggl({e^2 \over 4 \pi \epsilon_0 G}
\Biggr)^{\frac12} \; . \label{exact}
\end{equation}
Like the example of Section~\ref{intro} it is finite but not
analytic in the coupling constants $e^2$ and $G$. Unlike this
example, it diverges for small $G$. This is because gravity has
regulated the linear self-energy divergence which results for a
non-gravitating charged particle.

One can understand the process from the fact that gravity has a
built-in tendency to oppose divergences. A charge shell does not
want to contract in pure electromagnetism; the act of compressing it
calls forth a huge energy density concentrated in the nearby
electric field. Gravity, on the other hand, tends to make things
collapse, especially large concentrations of energy density. The
dynamical signature of this tendency is the large negative energy
density concentrated in the Newtonian gravitational potential. In
the limit the two effects balance and a finite total mass results.

Said this way, there seems no reason why gravitational interactions
should not act to cancel divergences in quantum field theory
\cite{cancel}. It is especially significant, in this context, that
the divergences of some quantum field theories --- such as QED ---
are weaker than the linear ones which ADM have shown that classical
gravity regulates. The frustrating thing is that one cannot hope to
see the cancellation perturbatively. In perturbation theory the
gravitational response to an effect at any order must be delayed to
a higher order. This is why the perturbative result (\ref{pert})
consists of an oscillating series of ever higher divergences. What
is needed is an approximation technique in which gravity knows what
is happening in the gauge sector so the gravitational response can
keep pace at the same order.

A final point of interest is that any finite bare mass drops out of
the exact result (\ref{exact}) in the limit $R \rightarrow 0$. This
makes for an interesting contrast with the usual program of
renormalization. Without gravity one would pick the desired physical
mass, $M_{\rm phys}$, and then adjust the bare mass to be whatever
divergent quantity was necessary to give it,
\begin{equation}
M_0 = M_{\rm phys} - {e^2 \over 8 \pi \epsilon_0 R c^2} \; .
\end{equation}
Of course the same procedure would work with gravity as well,
\begin{equation}
M_0 = M_{\rm phys} - {e^2 \over 8 \pi \epsilon_0 R c^2} + {G M_{\rm
phys}^2 \over 2 \epsilon_0 R c^2} \; .
\end{equation}
The difference with gravity is that we have an alternative: keep
$M_0$ finite and let the dynamical cancellation of divergences
produce a unique result for the physical mass. The ADM mechanism is
in fact the classical realization of the old dream of computing a
particle's mass from its self-interactions \cite{dream}.

\section{A New Expansion for Particle Masses}

The purpose of this section is to explain the new expansion we
propose for particle masses. For simplicity we work in the context
of a charged and gravitating scalar field, although the same
technique applies to fermions and to Yang-Mills force fields. The
Lagrangian is the sum of those for general relativity,
electrodynamics and a charged scalar,
\begin{eqnarray}
\mathcal{L}_{\rm GR}[g] & = & \frac{R \sqrt{-g}}{16 \pi G} \; ,
\label{LGR} \\
\mathcal{L}_{\rm EM}[g,A] & = & -\frac14 F_{\rho\sigma} F_{\mu\nu}
g^{\rho\mu} g^{\sigma\nu} \sqrt{-g} \; , \label{LEM} \\
\mathcal{L}_{\rm SC}[g,A,\phi,\phi^*] & = & - (D_{\mu} \phi)^*
D_{\nu} \phi g^{\mu\nu} \sqrt{-g} - M_0^2 \phi^* \phi \sqrt{-g} \; .
\qquad \label{LSC}
\end{eqnarray}
Here and henceforth $g_{\mu\nu}(t,\vec{x})$ stands for the metric
field, with inverse $g^{\mu\nu}$ and determinant $g$;
$A_{\mu}(t,\vec{x})$ denotes the electromagnetic vector potential
with field strength $F_{\mu\nu} \equiv \partial_{\mu} A_{\nu} -
\partial_{\nu} A_{\mu}$; and $\phi(t,\vec{x})$ is the complex scalar
field. The covariant derivative operator is $D_{\mu} \equiv
\partial_{\mu} + i e A_{\mu}$. The alert reader will note that the
scalar Lagrangian lacks the quartic self-interaction that would be
required for perturbative renormalizability in flat space. Because
the charged scalar is anyway not perturbatively renormalizable once
gravity has been included, there does not seem to be any point to
including this term for a first investigation of nonperturbative
renormalizability.

We employ the usual units of particle physics in which $\hbar = 1 =
c$, so that time and space have the dimensions of inverse mass, the
charge $e$ is a pure number, the Newton constant $G$ is an inverse
mass-squared, and the bare mass $M_0$ is a mass. We shall also
sometimes distinguish time and space arguments --- as in
$\phi(t,\vec{x})$ --- and sometimes lump them together into a single
spacetime coordinate $x^{\mu} = (t,\vec{x})$ --- as in $\phi(x)$.

Our attitude is that the physical mass $M$ of single scalar states
is some complicated function of the bare parameters, $e$, $G$ and
$M_0$. Our first goal is to derive a formal expression that would
give $M$, assuming we had infinite computational ability. We then
develop an alternative to the conventional perturbative expansion
for evaluating this formal expression.

\subsection{Functional Integral Expression for the Mass}

If all interactions were turned off it would be simple to express
the free scalar field in terms of the operators
$b^{\dagger}(\vec{k})$ and $a(\vec{k})$ which create and annihilate
one particle states with wave number $\vec{k}$,
\begin{equation}
\phi_{\rm free}(t,\vec{x}) = \int \!\! \frac{d^3 k}{(2\pi)^3}
\Biggl\{ \frac1{\sqrt{2 \omega_0}} \, e^{-i\omega_0 t + i \vec{k}
\cdot \vec{x}} a(\vec{k}) + \frac1{\sqrt{2 \omega_0}} \, e^{i \omega_0
t - \vec{k} \cdot \vec{x}} b^{\dagger}(\vec{k}) \Biggr\} \; .
\label{freefield}
\end{equation}
Here $\omega_0 \equiv \sqrt{k^2 + M_0^2}$ is the free energy. We can
invert relation (\ref{freefield}) to solve for the annihilation
operator using the Wronskian $\overleftrightarrow{W}_{\!\!\mu} \equiv
\overleftarrow{\partial}_{\!\!\mu} - \overrightarrow{\partial}_{\!\!\mu}$,
\begin{equation}
a(\vec{k}) = \frac{i}{\sqrt{\omega_0}} \, e^{i \omega_0 t} \,
\overleftrightarrow{W}_{\!\!0} \, \widetilde{\phi}_{\rm free}(t,\vec{k}) \; .
\label{freeann}
\end{equation}
Here and henceforth, a tilde over some function denotes its spatial
Fourier transform,
\begin{equation}
\widetilde{f}(t,\vec{k}) \equiv \int \!\! d^3x \, e^{-i \vec{k}
\cdot \vec{x}} f(t,\vec{x}) \; .
\end{equation}

In the presence of interactions it is no longer possible to give
explicit relations such as (\ref{freeann}) for the operators which
create and destroy exact 1-particle states. However, if we
temporarily regulate infrared divergences and agree to understand
operator relations in the weak sense then it is possible to write
the operators which annihilate outgoing particles and create
incoming ones as simple limits \cite{LSZ},
\begin{eqnarray}
a^{\rm out}(\vec{k}) & = & \lim_{t_{+} \rightarrow \infty} \frac{i
e^{i \omega t_{+}}}{\sqrt{2 \omega Z}} \, \overleftrightarrow{W}_{\!\!t_{+}}
\, \widetilde{\phi}(t_+,{\vec k}) \; , \label{asymp1} \\
\Bigl(a^{\rm in}(\vec{k})\Bigr)^{\dagger} & = & \lim_{t_{-}
\rightarrow -\infty} \frac{i e^{-i\omega t_{-}}}{\sqrt{2 \omega Z}} \,
\overleftrightarrow{W}_{\!\!t_{-}} \, \widetilde{\phi}^*(t_-,{\vec k}) \; .
\label{asymp2}
\end{eqnarray}
Here $Z$ is the field strength renormalization (defined as the
amplitude for the field to create a 1-particle state) and $\omega$
is the full energy,
\begin{equation}
\omega \equiv \sqrt{k^2 + M^2} \; . \label{omega}
\end{equation}
At this stage we do not know the physical mass $M$; it is some
function of the bare parameters of the theory.

Now consider single particle states whose wave functions in the
infinite past and future are $\psi_{\mp}(\vec{k})$, respectively. We
can employ relations (\ref{asymp1}-\ref{asymp2}) to derive an
expression for the inner product between these states,
\begin{eqnarray}
\lefteqn{ \Bigl\langle \psi_{+}^{\rm out} \Bigl\vert \psi_{-}^{\rm
in} \Bigr\rangle = \int \!\! \frac{d^3k}{(2\pi)^3} \, \frac1{2 Z
\omega} \psi_{+}^*(\vec{k}) \psi_{-}(\vec{k}) \Bigl[\lim_{t_{+}
\rightarrow \infty} e^{i \omega t_{+}}
\overleftrightarrow{W}_{\!\!t_{+}} \bullet \Bigr] \Bigl[\lim_{t_{-}
\rightarrow \infty} e^{-i \omega t_{-}}
\overleftrightarrow{W}_{\!\!t_{-}} \bullet \Bigr] } \nonumber \\
& & \hspace{4.5cm} \times \int \!\! d^3x \, e^{-i \vec{k} \cdot
\vec{x}} \Bigl\langle \Omega^{\rm out} \Bigr\vert
\phi(t_{+},\vec{x}) \phi^*(t_{-},\vec{0}) \Bigr\vert \Omega^{\rm in}
\Bigr\rangle \; . \qquad \label{complicated}
\end{eqnarray}
One way of computing the physical mass $M$ would be to adjust it to
the precise value for which expression (\ref{complicated}) reduces
to,
\begin{equation}
\Bigl\langle \psi_{+}^{\rm out} \Bigl\vert \psi_{-}^{\rm in}
\Bigr\rangle = \int \!\! \frac{d^3k}{(2\pi)^3} \, \frac1{2 Z \omega}
\, \psi_{+}^*(\vec{k}) \psi_{-}(\vec{k}) \; .
\end{equation}
This agrees with the usual definition of the mass as the pole of the
propagator, but it is problematic owing to infrared divergences.

A more direct way of computing the mass is to focus on the second
line of (\ref{complicated}) which we can express as the exponent of
$-i$ times some complex function $\xi(t_{+},t_{-},k)$,
\begin{equation}
e^{-i \xi(t_{+},t_{-},k)} \equiv \int \!\! d^3x \, e^{-i \vec{k}
\cdot \vec{x}} \, \Bigl\langle \Omega^{\rm out} \Bigl\vert
\phi(t_{+},\vec{x}) \phi^*(t_{-},\vec{0}) \Bigr\vert \Omega^{\rm
in} \Bigr\rangle \; . \label{xidef}
\end{equation}
This function $\xi(t_{+},t_{-},k)$ includes many things --- the
field strength renormalization, finite time correlation effects from
multiparticle states, and so on --- but only the single particle
energy grows linearly with the time interval. Dividing by the time
interval and then taking it to infinity gives this energy,
\begin{equation}
\lim_{t_{\pm} \rightarrow \pm\infty} \Biggl[ \frac{
\xi(t_{+},t_{-},k)}{t_{+} \!-\! t_{-}} \Biggr] = \sqrt{k^2 + M^2} \;
.
\end{equation}
Setting $k = 0$ gives the physical mass we are seeking. This way of
computing $M$ avoids the problems of infrared divergences which
affect the field strength renormalization but not the mass.

It is straightforward to write (\ref{xidef}) as a functional
integral,
\begin{eqnarray}
\lefteqn{e^{-i \xi(t_{+},t_{-},k)} = \int \!\! d^3x \, e^{-i \vec{k}
\cdot \vec{x}} } \nonumber \\
& & \hspace{.4cm} \times \Fint [dg] [dA] [d\phi] [d\phi^*] \,
\phi(t_{+},\vec{x}) \phi^*(t_{-},\vec{0}) \, e^{iS_{\rm GR}[g] + i
S_{\rm EM}[g,A] + i S_{\rm SC}[g,A,\phi,\phi^*]} \; . \qquad
\label{funcint}
\end{eqnarray}
We have subsumed the details of gravitational and electromagnetic
gauge fixing into the measure factors $[dg]$ and $[dA]$. Note that
the various action integrals in expression (\ref{funcint}) go from
time $t_{-}$ to $t_{+}$,
\begin{eqnarray}
S_{\rm GR}[g] & \equiv & \int_{t_{-}}^{t_{+}} \!\! dt \! \int \!\! d^3x \,
\mathcal{L}_{\rm GR}[g](t,\vec{x}) \; , \qquad \\
S_{\rm EM}[g,A] & \equiv & \int_{t_{-}}^{t_{+}} \!\! dt \! \int \!\! d^3x \,
\mathcal{L}_{\rm EM}[g,A](t,\vec{x}) \; , \qquad \\
S_{\rm SC}[g,A,\phi,\phi^*] & \equiv & \int_{t_{-}}^{t_{+}} \!\! dt \! \int
\!\! d^3x \, \mathcal{L}_{\rm SC}[g,A,\phi,\phi^*](t,\vec{x}) \; . \qquad
\end{eqnarray}

\subsection{Eliminating the Matter Field}

Expression (\ref{funcint}) is only formal because there is not yet
any way of defining it or evaluating it. One would usually resort
to perturbation theory at this point. We shall instead integrate
out the scalar field,
\begin{eqnarray}
\lefteqn{e^{-i\xi(t_{+},t_{-},k)} = \int \!\! d^3x \, e^{-i \vec{k}
\cdot \vec{x}} } \nonumber \\
& & \hspace{2cm} \times \Fint [dg] [dA] \,
i\Delta[g,A]\Bigl(t_{+},\vec{x};t_{-},\vec{0}\Bigr) e^{iS_{\rm
GR}[g] + i S_{\rm EM}[g,A] + i \Gamma_{\rm SC}[g,A]} \; . \qquad
\label{newfint}
\end{eqnarray}
The two new quantities this produces are the scalar propagator
$i\Delta[g,A](x;x')$ and the scalar effective action $\Gamma_{\rm
SC}[g,A]$. Both can be defined using the scalar kinetic operator in
the presence of an arbitrary metric and vector potential,
\begin{equation}
\mathcal{D}[g,A] \equiv D_{\mu} \sqrt{-g} g^{\mu\nu} D_{\nu} - M_0^2
\sqrt{-g} \; .
\end{equation}
In rough terms, the scalar propagator is $i$ times the functional
inverse of $\mathcal{D}[g,A]$, subject to Feynman boundary
conditions, while the effective action is $i$ times the logarithm of
its determinant,
\begin{eqnarray}
\mathcal{D}[g,A] \times i\Delta[g,A](x;x') = i \delta^4(x \!-\! x')
\; , \label{fullprop} \\
\Gamma_{\rm SC}[g,A] \equiv i \ln\Bigl(\det[\mathcal{D}[g,A] \!-\! i
\epsilon]\Bigr) \; . \label{Gamma}
\end{eqnarray}

It will facilitate subsequent work to be more precise about the
definition of of the scalar propagator $i\Delta[g,A](x;x')$.
Consider a general ``mode function'' $u[g,A](t,\vec{x};\lambda)$
which obeys the homogeneous equation,
\begin{equation}
\mathcal{D}[g,A] \times u[g,A](x;\lambda) = 0 \; . \label{KGeqn}
\end{equation}
Here ``$\lambda$'' is a (possibly continuous) index which labels the
solution; it would be the wave vector $\vec{k}$ for flat space and
zero potential. We additionally require that the set of all
solutions obey the canonical normalization condition,
\begin{equation}
-i \int_{t = {\rm const}} \!\!\!\!\!\!\!\!\!\!\!\! d^3x \sqrt{-g(x)}
\, g^{0\nu}(x) \, u[g,A](x;\lambda)
\Bigl(\overleftarrow{D}_{\!\!\nu} - \overrightarrow{D}_{\!\!\nu}^*
\Bigr) u^*[g,A](x;\kappa) = \delta_{\lambda\kappa} \; .
\label{normalization}
\end{equation}
In terms of these mode functions the propagator is,
\begin{eqnarray}
\lefteqn{i\Delta[g,A](x;x') = \sum_{\lambda} \Biggl[ \theta(t \!-\!
t') u[g,A](x;\lambda) u^*[g,A](x';\lambda) } \nonumber \\
& & \hspace{5cm} + \theta(t' \!-\! t) u^*[g,A](x;\lambda)
u[g,A](x';\lambda) \Biggr] \; . \qquad
\end{eqnarray}
Note that only the first theta function contributes in the limit we
require,
\begin{equation}
\lim_{t_{+} \gg t_{-}}
i\Delta[g,A]\Bigl(t_{+},\vec{x};t_{-},\vec{0}\Bigr) = \sum_{\lambda}
u[g,A](t_{+},\vec{x};\lambda) u^*[g,A](t_{-},\vec{0};\lambda) \; .
\end{equation}
Our expression for the physical mass can therefore be written as,
\begin{eqnarray}
\lefteqn{M = \lim_{t_{\pm} \rightarrow \pm \infty} \Biggl(
\frac{i}{t_{+} \!-\! t_{-}} \Biggr) \ln\Biggl[ \int \!\! d^3x
\sum_{\lambda} \Fint [dg] [dA] } \nonumber \\
& & \hspace{1cm} \times u[g,A](t_{+},\vec{x};\lambda) \,
u^*[g,A](t_{-},\vec{0};\lambda) \, e^{i S_{\rm GR}[g] + i S_{\rm
EM}[g,A] + i \Gamma_{\rm SC}[g,A]} \Biggr] \; . \qquad
\label{finalform}
\end{eqnarray}

\subsection{Stationary Phase Expansion}

The expression (\ref{finalform}) we have derived for the physical
mass is exact, but formal because no one knows how to evaluate the
functional integral. That would ordinarily be done by resorting to
conventional perturbation theory. This could be viewed as an
application of a simplified variant of the method of stationary
phase to the original functional integral (\ref{funcint}). Because
our modified expansion involves undoing some of the simplifications
we digress to review the technique.

Recall that the method of stationary phase gives an asymptotic
expansion for integrals of the form,
\begin{equation}
I \equiv \int \!\! dz \, e^{i f(z)} \; .
\end{equation}
The technique is to first find the stationary point $z_0$ (which we
assume to be unique) such that $f'(z_0) = 0$. One then expands
$f(z)$ around $z_0$,
\begin{eqnarray}
f(z) & = & f(z_0) + \frac12 f''(z_0) (z \!-\! z_0)^2 +
\sum_{n=3}^{\infty} \frac{f^{(n)}(z_0)}{n!} \, (z \!-\! z_0)^n \; ,
\qquad \\
& \equiv & f_0 + \frac12 f_0'' (z \!-\! z_0)^2 + \Delta \!f(z \!-\!
z_0) \; . \qquad
\end{eqnarray}
The next step is shifting to the variable $\zeta \equiv z - z_0$ and
expanding $e^{i\Delta \!f}$,
\begin{equation}
I = e^{if_0} \int \!\! d\zeta \, e^{\frac{i}2 f_0'' \zeta^2}
\sum_{m=0}^{\infty} \frac1{m!} \Bigl[ i \Delta \!f(\zeta)\Bigr]^m \;
.
\end{equation}
At this stage the result is still exact, but generally no simpler to
evaluate than the original form. What gives a computable series is
the final step of interchanging integration and summation. It is at
this point that the expansion ceases to be exact and becomes only
asymptotic,
\begin{eqnarray}
I & \longrightarrow & e^{if_0} \sum_{m=0}^{\infty} \frac{i^m}{m!}
\int \!\! d\zeta \, e^{\frac{i}2 f_0'' \zeta^2} \, \Bigl[ \Delta
\!f(\zeta)\Bigr]^m \; , \\
& = & e^{i f_0} \times \frac{e^{\frac{i}4 \pi}}{\sqrt{2 \pi f_0''}}
\Biggl[1 + \frac18 \times i f_0'''' \Bigl(\frac{i}{f_0''}\Bigr)^2 +
\frac{5}{24} \times (i f_0''')^2 \Bigl(\frac{i}{f_0''}\Bigr)^3 +
\cdots \Biggr] \; . \qquad
\end{eqnarray}

Conventional perturbation theory is a simplified form of this
technique, applied to the functional integral (\ref{funcint}). The
two simplifications are:
\begin{itemize}
\item{The multiplicative factor of $\phi(t_{+},\vec{x})
\phi^*(t_{-},\vec{0})$ is not included in the exponential, along
with the action; and}
\item{The stationary field configuration is assumed to be flat space
with no charge fields,
\begin{equation}
\overline{g}_{\mu\nu}(x) = \eta_{\mu\nu} \quad , \quad
\overline{A}_{\mu}(x) = 0 \quad , \quad \overline{\phi}(x) = 0 \; .
\end{equation}}
\end{itemize}
These two simplification make conventional perturbation theory much
simpler and more generally applicable than a strict application of
the method of stationary phase because they remove any dependence of
the propagators and vertices on the operator whose expectation value
is being computed --- in this case $\phi(t_{+},\vec{x})
\phi^*(t_{-},\vec{0})$. However, simplicity is not always desirable,
nor is it always physically correct. In this case, those
simplifications preordain that the eventual asymptotic expansion for
$M$ can contain only integer powers of $G$ and $e^2$. The same two
simplifications also imply that the gravitational response to an
$\ell$ loop effect in the electromagnetic sector must be delayed
until $\ell+1$ loop order.

Our modified expansion is defined by making three changes. The first
is to integrate out the matter fields and start from the functional
integral (\ref{finalform}). That is not so important. The important
change is the second one: we include the factor
$u[g,A](t_{+},\vec{x};\lambda) \times
u^*[g,A](t_{-},\vec{0};\lambda)$ in the exponential, along with the
gravitational and electromagnetic actions \cite{TW1}. This makes the
eventual series vastly more complicated but it does three desirable
things:
\begin{itemize}
\item{It allows the apparatus of perturbation theory --- propagators and
vertices --- to depend on the operator whose expectation value is
being computed;}
\item{It permits the eventual expansion to involve complicated,
nonanalytic functions of $e^2$ and $G$; and}
\item{It allows the gravitational response to an electromagnetic effect
at some order to occur at the same perturbative order.}
\end{itemize}

The final change is that we drop the scalar effective action
$\Gamma_{\rm SC}[g,A]$ from how we define the expansion. That is, it
plays no role in determining the stationary point, the ``classical''
action, the propagators or the vertices; we regard $e^{i \Gamma_{\rm
SC}}$ as a multiplicative factor, like $\phi \phi^*$ was in the
conventional expansion. Although this can be done consistently
because $\Gamma_{\rm SC}[g,A]$ is gauge invariant, there is no
physical justification for it. The scalar effective action
incorporates one loop vacuum polarization and its gravitational
analog, which may be important effects. Our justification is just
the pragmatic one that including $\Gamma_{\rm SC}[g,A]$ leads to a
more complicated set of equations for the stationary point than we
presently understand how to solve.

\subsection{Physical Interpretation}

The 0th order term in the new expansion can be interpreted as the
energy of a first-quantized Klein-Gordon particle which moves in
the electromagnetic and gravitational fields that are sourced by
its probability current \cite{RPW2}. To see this, note that the
quantum mechanical propagator for such a Klein-Gordon particle
in fixed background fields $A_{\mu}(x)$ and $g_{\mu\nu}(x)$ is,
\begin{equation}
\mathcal{P}[g,A](x;x') = \sum_{\lambda} u[g,A](x;\lambda) \,
u^*[g,A](x';\lambda) \; .
\end{equation}
That means we can evolve the first-quantized wave function from
$t'$ to $t$ by taking the inner product with $\mathcal{P}[g,A](x;x')$
according to relation (\ref{normalization}),
\begin{equation}
\int_{t' = {\rm const}} \!\!\!\!\!\!\!\!\!\!\!\! d^3x' \sqrt{-g(x')}
\, g^{0\nu}(x') \, \mathcal{P}[g,A](x;x') \overleftrightarrow{W}_{\!\!\nu}'
\, u[g,A](x';\lambda) = u[g,A](x;\lambda) \; .
\end{equation}
In expression (\ref{finalform}) we are going from a delta function
$\psi_{-}(\vec{x}') = \delta^3(\vec{x}')$ to a zero momentum plane
wave $\psi_{+}(\vec{x}) = 1$. Of course the stationary field
configurations for the metric and the vector potential are just
those sourced by the quantum mechanical particle itself, hence the
stated interpretation of the 0th order term.

Our ability to evaluate the 0th order term depends upon whether or
not the first-quantized particle can form a bound state in its own
potentials. If not then we are left with a complicated scattering
problem which seems to be intractable. However, many
simplifications are possible if a bound state forms. First, one can
forget about the continuum solutions; the result for $M$ in expression
(\ref{finalform}) will derive entirely from the bound state with the
largest overlap between the two asymptotic wavefunctions $\psi_{\pm}$.
Second, one can specialize the wave function to take the form,
\begin{equation}
u[\overline{g},\overline{A}](t,\vec{x}) = e^{-i E t} F(r) \;
\label{ansatz1}
\end{equation}
Third, in solving for the stationary potentials one need only consider
a class of metrics and vector potentials which is broad enough to include
the eventual solution. For scalar QED this reduces the potentials from
nine functions of spacetime (after gauge fixing) to only three functions
of a single variable,
\begin{eqnarray}
\overline{g}_{\mu\nu}(x) dx^{\mu} dx^{\nu} & = & -B(r) dt^2
+ A(r) dr^2 + r^2 d\Omega^2 \; , \qquad \label{ansatz2} \\
\overline{A}_{\mu}(x) dx^{\mu} & = & -\Phi(r) dt \; . \qquad
\label{ansatz3}
\end{eqnarray}
This simplification might also be relevant to the problem of going
beyond 0th order. Although it is impossible to work out propagators
for general background fields, it is sometimes possible to derive
the propagator for potentials which depend upon only a single
variable \cite{TW2}.

A final simplification is the possibility of deriving a variational
formula for the bound state energy. Even if the functions $F(r)$,
$A(r)$, $B(r)$ and $\Phi(r)$ could not be determined exactly, this
technique would allow us to explore reasonable approximations for
them. We might even optimize the free parameters in the largest
class of solutions for which the field propagators can be worked out.
A variational technique would also give an upper bound on the bound
state energy.

The reason deriving a variational technique is only a possibility
is that this system involves gravitation. It is well known that not
all of the gravitational potentials contribute positive energy
\cite{ADM2}. Of course it is precisely the negative energy potentials
that provide the possibility for canceling self-energy divergences!
These negative energy potentials do not lead to any instability
because they are completely constrained variables; that is, they are
determined in terms of the other variables. However, it is the
constraint equations which enforces these relations. The Hamiltonian
is not bounded below before imposing these constraint equations.
So being able to derive a variational formalism depends upon
identifying the negative energy potentials and solving the
constraint equations for them.

\section{Rearrangement of Perturbation Theory}

The purpose of this section is to show where the old diagrams end up
in the new expansion. The point of doing this is not to perform the
computation; there are more efficient techniques which will be
developed in the next section. The point is rather to see that the
new expansion offers gravity the chance of ``keeping up'' with what
is happening in other sectors. We will show that all the usual
diagrams are present, and at the same ``loop'' order as in
conventional perturbation theory. However, they are joined with a
vast class of new diagrams which are added to one order and
subtracted from another. To see all this we resort to a very simple,
zero dimensional model of the functional integral in expression
(\ref{funcint}). We first work out what the conventional expansion
gives, then derive the zeroth and first order results for the new
expansion. The section closes with a discussion of the implications
for the new expansion.

\subsection{Zero Dimensional Model}

We can model the functional integral (\ref{funcint}) as an ordinary
integration over two $N$-vectors: $x_i$, representing the metric and
gauge fields; and $y^a$ representing the scalar. A simple model for
the action is,
\begin{equation}
S(\vec{x},\vec{y}) \equiv \frac12 A_{ij} x_i x_j + \frac12 B^{ab}
y^a y^b + \frac12 C^{ab}_i x_i y^a y^b \; . \label{0action}
\end{equation}
Of course scalar QED includes interactions between two photons and
two scalars, and the action of general relativity contains
self-interactions of the metric, as well as infinite order
interactions of the metric with the scalar and the vector potential.
However, our model action (\ref{0action}) will suffice for the
purpose of understanding how conventional perturbation theory is
reorganized.

\begin{figure}
\includegraphics[width=3cm,height=3cm]{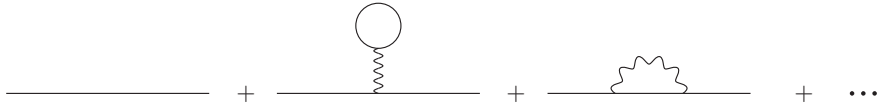}
\caption{Usual expansion for the scalar propagator. Solid lines are
scalars, wavy lines represent photons and gravitons.}
\label{fig1}
\end{figure}

The normalization integral and the 2-point function of our model
are,
\begin{eqnarray}
J & \equiv & \int \!\! d^N\!x \! \int \!\! d^N\!y \, e^{i S(\vec{x},\vec{y})}
\; , \label{Idef} \\
J^{ab} & \equiv & J^{-1} \int \!\! d^N\!x \! \int \!\! d^N\!y \, y^a y^b
e^{i S(\vec{x},\vec{y})} \; .
\end{eqnarray}
It will simplify the notation if we use the script letters
$\mathcal{A}_{ij}$ and $\mathcal{B}^{ab}$ to denote the matrix
inverses of $A_{ij}$ and $B^{ab}$,
\begin{equation}
A_{ij} \mathcal{A}_{jk} = \delta_{ik} \qquad , \qquad B^{ab}
\mathcal{B}^{bc} = \delta^{ac} \; .
\end{equation}
When multiplied by $i$ these inverses are the ``propagators'' of this
one dimensional quantum field theory. The expansions of the
normalization integral and the 2-point function are,
\begin{equation}
J = \frac{(2 \pi)^N e^{\frac{N \pi}{2} i } }{\sqrt{\det(A) \det(B) }}
\Biggl\{1 \!+\! \frac18 i C^{ab}_i i C^{cd}_j
i\mathcal{A}_{ij} \Bigl[ i\mathcal{B}^{ab} i\mathcal{B}^{cd} \!+\!
2 i \mathcal{B}^{ac} i\mathcal{B}^{bd}\Bigr] + O(C^4) \Biggr\} .
\end{equation}
\begin{equation}
J^{ab} = i\mathcal{B}^{ab} + \frac12 i\mathcal{B}^{ac} iC^{cd}_i
i\mathcal{A}_{ij} iC^{ef}_j i\mathcal{B}^{ef} i\mathcal{B}^{db} +
i\mathcal{B}^{ac} iC^{cd}_i i\mathcal{A}_{ij} i\mathcal{B}^{de} i
C^{ef}_j i\mathcal{B}^{fb} + O(C^4) \; . \label{2point}
\end{equation}
Fig.~\ref{fig1} gives the diagrams associated with the expansion
(\ref{2point}).

Following subsection 3.2, we integrate out the ``matter fields''
$y^a$,
\begin{equation}
\int \!\! d^N\!y \, e^{i S(\vec{x},\vec{y})} \, y^a y^b  =
e^{\frac{i}2 A_{ij} x_i x_j} \Biggl[ \frac{i}{B \!+\! C_k x_k}
\Biggr]^{ab} \frac{ (2\pi)^{\frac{N}2} e^{\frac{N \pi}4 i} }{
\sqrt{\det(B \!+\! C_{\ell} x_{\ell} )}} \; .
\end{equation}
At this stage it is useful to extract the lowest order contribution
from the determinant,
\begin{equation}
\det(B \!+\! C_{\ell} x_{\ell}) = \det(B) \times \det(I \!-\!
i\mathcal{B} \cdot i C_{\ell} x_{\ell}) \; .
\end{equation}
It is also useful to contract into ``asymptotic state
wavefunctions'' --- the analogues of $\psi_{\pm}$. These are the
$N$-vectors $u^a$, analogous to $\psi_{+}$, and $v^b$, analogous to
$\psi_{-}$. We shall also include a phase so that $u^a
i\mathcal{B}^{ab} v^b = 1$. Then the contraction of these two
vectors into the field dependent propagator, the analogue of
$i\Delta[g,A](x;x')$, can be written as,
\begin{eqnarray}
u^a \times \Biggl[ \frac{1}{B \!+\! C_k x_k}\Biggr]^{ab} \times v^b
& = & u^a \times \Biggl[ \frac{i}{I \!-\! i \mathcal{B} \cdot i C_k
x_k} \Biggr]^{ac} \times i\mathcal{B}^{cb} \times v^b \; , \qquad \\
& = & 1 + u^a \times \sum_{n=1}^{\infty} \Biggl[ \Bigl(i\mathcal{B}
\cdot i C_k x_k \Bigr)^n \Biggr]^{ac} \!\! \times
i\mathcal{B}^{cb} \times v^b \; , \qquad \\
& \equiv & 1 + i \Delta(\vec{x}) \; . \qquad
\end{eqnarray}
Except for the phase, the quantity whose logarithm we wish to take
is,
\begin{equation}
\mathcal{J} = u^a \times J^{ab} \times v^b
= \frac{ \sqrt{\det(A)}}{ (2\pi)^{\frac{N}2} e^{\frac{N \pi}4
i}} \int \!\! d^N\!x \, e^{i \mathcal{E}(\vec{x}) + i
\Gamma(\vec{x})} \; . \label{Jay}
\end{equation}
Here the exponent $\mathcal{E}(\vec{x})$ and the ``matter effective
action'' $\Gamma(\vec{x})$ are,
\begin{equation}
\mathcal{E}(\vec{x}) = \frac12 A_{ij} x_i x_j \!-\! i \ln\Bigl[1
\!+\! i \Delta(\vec{x}) \Bigr] \quad , \quad
\Gamma(\vec{x}) = \frac{i}2 \ln\Bigl[\det(I \!-\! i\mathcal{B}
\!\cdot\! i C_k x_k) \Bigr] \; .
\end{equation}

\subsection{0th Order in the Model}

\begin{figure}
\includegraphics[width=3cm,height=3cm]{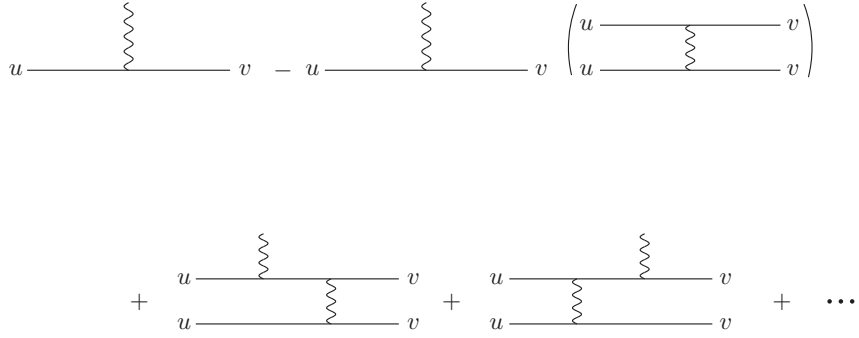}
\caption{Expansion for the solution $X_i$ to equation (\ref{XEOM}).
Solid lines are $y^a$ propagators (scalars), wavy lines represent
$x_i$ propagators (photons and gravitons).}
\label{fig2}
\end{figure}

We want to do the stationary phase expansion on expression
(\ref{Jay}), but treating $\Gamma(\vec{x})$ as higher order. Hence
the zeroth order term in the expansion is,
\begin{equation}
\mathcal{J}_0 \equiv e^{i \mathcal{E}(\vec{X})} = \Bigl[1 \!+\! i
\Delta(\vec{X})\Bigr] e^{\frac{i}2 A_{ij} X_i X_j} \; , \label{Jay0}
\end{equation}
where the ``stationary field configuration'' $X_i$ is found by
solving,
\begin{equation}
\frac{\partial \mathcal{E}(\vec{x})}{\partial x^i}
\Biggl\vert_{\vec{x} = \vec{X}} = A_{ij} X_j - \frac{i
\frac{\partial i \Delta(\vec{X})}{\partial x_i}}{1 \!+\! i
\Delta(\vec{X})} = 0 \; . \label{XEOM}
\end{equation}
Despite the simplicity of our model, equation (\ref{XEOM}) is too
difficult to solve exactly for general $A_{ij}$, $B^{ab}$ and $C^{ab}_i$.
However, we are only interested in a perturbative solution --- in
powers of the ``interaction'' $C^{ab}_i$ --- and that is simple enough
to generate by iteration,
\begin{eqnarray}
\lefteqn{X_i = \frac{i \mathcal{A}_{ij} \frac{\partial i\Delta(\vec{X})}{
\partial x_j}}{1 \!+\! i\Delta(\vec{X})} \; , } \\
& & \hspace{-.3cm} = i\mathcal{A}_{ij} u \Bigl[i\mathcal{B} i C_j
i\mathcal{B}\Bigr] v - i\mathcal{A}_{ij} u \Bigl[ i\mathcal{B} i C_j
i\mathcal{B}\Bigr] v \times u \Bigl[ i\mathcal{B} i C_k i\mathcal{B}
\Bigr] v i\mathcal{A}_{k\ell} u \Bigl[ i\mathcal{B} i C_{\ell}
i\mathcal{B} \Bigr] v \nonumber \\
& & \hspace{.3cm} + i\mathcal{A}_{ij} u \Bigl[i\mathcal{B} \Bigl(i C_j
i\mathcal{B} i C_k  \!+\! i C_k i\mathcal{B} i C_j \Bigr) i\mathcal{B}
\Bigr] v i\mathcal{A}_{k\ell} u \Bigl[ i\mathcal{B} i C_{\ell}
i\mathcal{B} \Bigr] v + O(C^5) \; . \qquad \label{Xexp}
\end{eqnarray}
Fig.~\ref{fig2} gives a diagrammatic representation of the expansion.

\begin{figure}
\includegraphics[width=2.75cm,height=2.75cm]{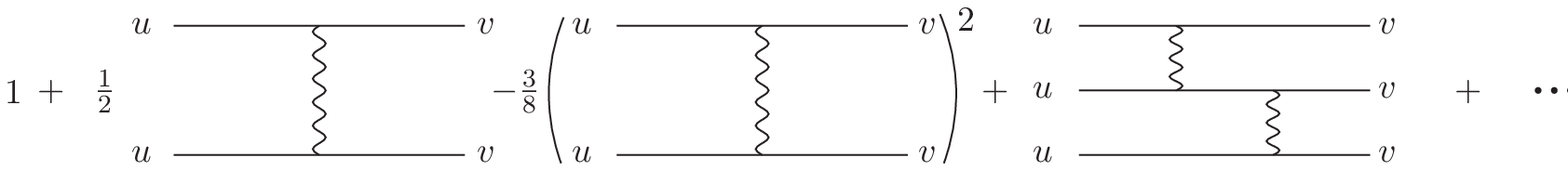}
\caption{Expansion for the 0th order term $\mathcal{J}_0$.
Solid lines are $y^a$ propagators (scalars), wavy lines represent
$x_i$ propagators (photons and gravitons).}
\label{fig3}
\end{figure}

In evaluating $\mathcal{J}_0$ it is useful to first formally express
things in terms of $i\Delta(\vec{X})$ and $\frac{i\Delta(\vec{X})}{
\partial x_i} X_i$,
\begin{eqnarray}
i\Delta(\vec{X}) & \!\!=\!\! & u \Bigl[ i\mathcal{B} \!\cdot\! iC_k X_k
\!\cdot\! i\mathcal{B} \Bigr] v + u \Bigl[ i\mathcal{B} \!\cdot\! iC_k X_k
\!\cdot\!  i\mathcal{B} \!\cdot\! iC_{\ell} X_{\ell} \!\cdot\! i\mathcal{B}
\Bigr] v + \cdots \; , \qquad \\
\frac{\partial i\Delta(\vec{X})}{\partial x_i} X_i & \!\!=\!\! & u \Bigl[
i\mathcal{B} \!\cdot\! iC_k X_k \!\cdot\! i\mathcal{B} \Bigr] v +
2 u \Bigl[ i\mathcal{B} \!\cdot\! iC_k X_k \!\cdot\! i\mathcal{B}
\!\cdot\! iC_{\ell} X_{\ell} \!\cdot\! i\mathcal{B} \Bigr] v +
\cdots \; . \qquad
\end{eqnarray}
Using expressions (\ref{Jay0}) and (\ref{XEOM}) we can write
$\mathcal{J}_0$ as,
\begin{eqnarray}
\mathcal{J}_0 & = & 1 + i\Delta(\vec{X}) -\frac12 \frac{\partial
i\Delta(\vec{x})}{\partial x_i} X_i + \frac{ [\frac{\partial i
\Delta(\vec{X})}{\partial x_i} X_i]^2 }{8 [1 \!+\! i \Delta(\vec{X})]}
+ \cdots \; , \qquad \\
& = & 1 + \frac12 u \Bigl[ i\mathcal{B} \!\cdot\! i C_k X_k \!\cdot\!
i\mathcal{B} \Bigr] v  + \frac18 \Biggl( u \Bigl[ i\mathcal{B} \!\cdot\!
i C_k X_k \!\cdot\! i\mathcal{B} \Bigr] v \Biggr)^2 + \dots \qquad
\label{penultimate}
\end{eqnarray}
Now just substitute (\ref{Xexp}) in (\ref{penultimate}) to obtain,
\begin{eqnarray}
\lefteqn{\mathcal{J}_0 = 1 + \frac12 u \Bigl[ i\mathcal{B} \!\cdot\!
i C_i \!\cdot\! i\mathcal{B} \Bigr] v \, i\! \mathcal{A}_{ij} \, u \Bigl[
i\mathcal{B} \!\cdot\! i C_j \!\cdot\! i\mathcal{B} \Bigr] v } \nonumber \\
& & \hspace{.5cm} -\frac38 \Biggl( u \Bigl[ i\mathcal{B} \!\cdot\!
i C_i \!\cdot\! i\mathcal{B} \Bigr] v \, i \! \mathcal{A}_{ij} \, u \Bigl[
i\mathcal{B} \!\cdot\! i C_j \!\cdot\! i\mathcal{B} \Bigr] v \Biggr)^2
+ u \Bigl[ i\mathcal{B} \!\cdot\! i C_i \!\cdot\! i\mathcal{B} \Bigr] v
\nonumber \\
& & \hspace{1.8cm} \times i \! \mathcal{A}_{ij} \,
u \Bigl[ i\mathcal{B} \!\cdot\! i C_j \!\cdot\!
i\mathcal{B} \!\cdot\! i C_k \!\cdot\! i\mathcal{B} \Bigr] v \, i\!
\mathcal{A}_{k\ell} \, u \Bigl[ i\mathcal{B} \!\cdot\! i C_{\ell} \!\cdot\!
i\mathcal{B} \Bigr] v + O(C^6) \; . \qquad \label{finalexp}
\end{eqnarray}
Fig.~\ref{fig3} gives a diagrammatic expansion of $\mathcal{J}_0$.

\subsection{1st Order in the Model}

One loop effects come from the determinantal correction to the
Method of Stationary Phase,
\begin{equation}
\mathcal{J}_{0+1} \equiv e^{i \mathcal{E}(\vec{X})} \times
\Biggl(\det\Biggl[ \mathcal{A}_{ij} \frac{\partial^2
\mathcal{E}(\vec{X})}{\partial x_j \partial x_k } \Biggr]
\Biggr)^{-\frac12} \; . \label{Jay01}
\end{equation}
The matrix whose determinant we must compute is,
\begin{equation}
\mathcal{A}_{ij} \frac{\partial^2 \mathcal{E}(\vec{X})}{\partial x_j
\partial x_k} = \delta_{ik} + \Biggl[ \frac{ X_i \frac{\partial
i \Delta(\vec{X})}{\partial x_k} \!-\! i\! \mathcal{A}_{ij}
\frac{\partial^2 i\Delta(\vec{X})}{\partial x_j \partial x_k}}{1
\!+\! i \Delta(\vec{X})} \Biggr] \equiv \delta_{ik} + M_{ik} \; .
\end{equation}
We can evaluate the determinant in (\ref{Jay01}) by first expanding
in powers of the matrix $M_{ij}$,
\begin{eqnarray}
\frac1{\sqrt{\det(I \!+\! M)}} & = & e^{-\frac12 {\rm Tr}[ \ln(I
+ M)]} \; , \\
& = & 1 -\frac12 {\rm Tr}[M] + \frac18 \Bigl( {\rm Tr}[M]\Bigr)^2
+ \frac14 {\rm Tr}[M^2] + O(M^3) \; . \qquad \label{detexp}
\end{eqnarray}

The determinantal contribution to (\ref{Jay01}) is sufficiently
complex that it is worthwhile to present results for each term of
(\ref{detexp}). As before, it is best to express these contributions
in terms of the full solution $X_i$,
\begin{eqnarray}
\lefteqn{-\frac12 {\rm Tr}[M] = -\frac12 u \Bigl[ i\mathcal{B} \!\cdot\!
i C_k X_k \!\cdot\! i\mathcal{B} \Bigr] v + i\! \mathcal{A}_{ij}
u \Bigl[ i\mathcal{B} \!\cdot\! i C_i \!\cdot\! i\mathcal{B} \!\cdot\!
i C_j \!\cdot\! i\mathcal{B} \Bigr] v } \nonumber \\
& & \hspace{-.3cm} +\frac12 \Biggl( u \Bigl[ i\mathcal{B} \!\cdot\!
i C_k X_k \!\cdot\! i\mathcal{B} \Bigr] v \Biggr)^{\!2} \!\!\!- u
\Bigl[ i\mathcal{B} \!\cdot\! i C_k X_k \!\cdot\! i\mathcal{B} \Bigr] v
\, i\! \mathcal{A}_{ij} \, u \Bigl[ i\mathcal{B} \!\cdot\! i C_i \!\cdot\!
i\mathcal{B} \!\cdot\! i C_j \!\cdot\! i\mathcal{B} \Bigr] v \nonumber \\
& & \hspace{.3cm} - u \Bigl[ i\mathcal{B} \!\cdot\! i C_k X_k \!\cdot\!
i\mathcal{B} \!\cdot\! i C_{\ell} X_{\ell} \!\cdot\! i\mathcal{B} \Bigr] v
+ i\! \mathcal{A}_{ij} u \Biggl[ i\mathcal{B} \Biggl( i C_i \!\cdot\!
i\mathcal{B} \!\cdot\! i C_j \!\cdot\! i\mathcal{B} \!\cdot\! i C_k X_k
\qquad \nonumber \\
& & \hspace{.7cm} + i C_i \!\cdot\! i\mathcal{B} \!\cdot\! i C_k X_k
\!\cdot\! i\mathcal{B} \!\cdot\! i C_j \!+\! i C_k X_k \!\cdot\!
i\mathcal{B} \!\cdot\! i C_i \!\cdot\! i\mathcal{B} \!\cdot\! i C_j
\Biggr) i\mathcal{B} \Biggr] v + O(C^6) \; , \qquad \\
\lefteqn{ \frac18 \Bigl( {\rm Tr}[M] \Bigr)^2 = \frac18 \Biggl( u \Bigl[
i\mathcal{B} \!\cdot\! i C_k X_k \!\cdot\! i\mathcal{B} \Bigr] v
\Biggr)^{\!2} -\frac12 u \Bigl[ i\mathcal{B} \!\cdot\! i C_k X_k \!\cdot\!
i\mathcal{B} \Bigr] v \, i\! \mathcal{A}_{ij} } \nonumber \\
& & \hspace{-.2cm} \times u \Bigl[ i\mathcal{B} \!\cdot\! i C_i \!\cdot\!
i\mathcal{B} \!\cdot\! i C_j \!\cdot\! i\mathcal{B} \Bigr] v + \frac12
\Biggl( i\! \mathcal{A}_{ij} \, u \Bigl[ i\mathcal{B} \!\cdot\! i C_i
\!\cdot\! i\mathcal{B} \!\cdot\! i C_j \!\cdot\! i\mathcal{B} \Bigr] v
\Biggr)^{\!2} + O(C^6) \; , \qquad \\
\lefteqn{ \frac14 {\rm Tr}[M^2] = \frac14 \Biggl( u \Bigl[ i\mathcal{B}
\!\cdot\! i C_k X_k \!\cdot\! i\mathcal{B} \Bigr] v \Biggr)^{\!2} \!\!-
u \Bigl[ i\mathcal{B} \!\cdot\! i C_k X_k \!\cdot\! i\mathcal{B} \!\cdot\!
i C_{\ell} X_{\ell} \!\cdot\! i\mathcal{B} \Bigr] v \!+\! \frac12 i\!
\mathcal{A}_{ij} \, i\! \mathcal{A}_{k\ell} } \nonumber \\
& & \hspace{-.3cm} \times u \Bigl[ i\mathcal{B} \!\cdot\! i C_j \!\cdot\!
i\mathcal{B} \!\cdot\! i C_k \!\cdot\! i\mathcal{B} \Bigr] v \,
u \Bigl[ i\mathcal{B} \Bigl( i C_{\ell} \!\cdot\! i\mathcal{B} \!\cdot\!
i C_i \!+\! i C_i \!\cdot\! i\mathcal{B} \!\cdot\! i C_{\ell} \Bigr)
i\mathcal{B} \Bigr] v + O(C^6) \; . \qquad
\end{eqnarray}

\begin{figure}
\includegraphics[width=5cm,height=6cm]{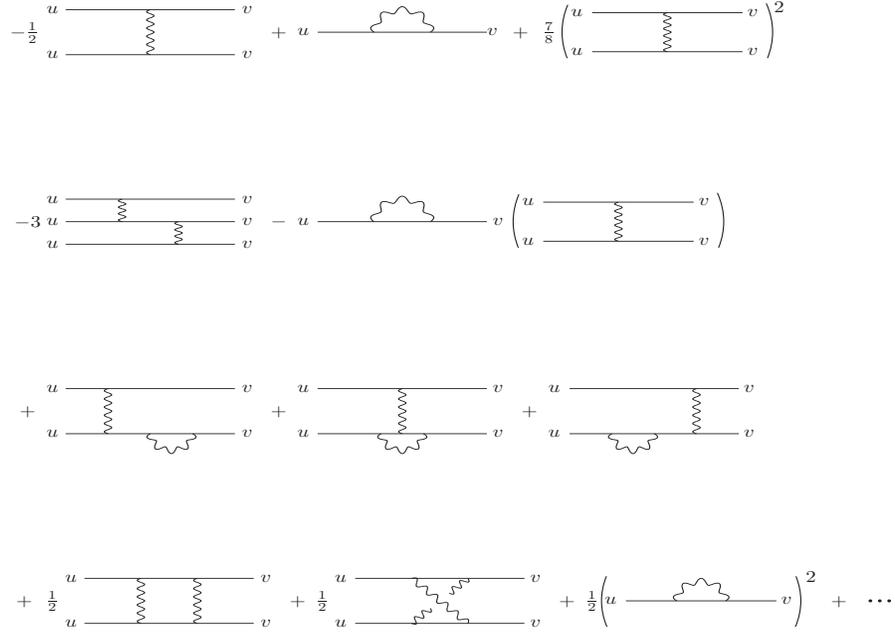}
\caption{Expansion for the 1st order term $\mathcal{J}_1$.
Solid lines are $y^a$ propagators (scalars), wavy lines represent
$x_i$ propagators (photons and gravitons).}
\label{fig4}
\end{figure}

The first order contribution to $\mathcal{J}$ in our new expansion is,
\begin{eqnarray}
\lefteqn{ \mathcal{J}_1 = e^{i \mathcal{E}(\vec{X})} \Biggl\{
\Biggl(\det\Biggl[ \mathcal{A}_{ij} \frac{\partial^2
\mathcal{E}(\vec{X})}{\partial x_j \partial x_k } \Biggr]
\Biggr)^{-\frac12} - 1 \Biggr\} \; , } \\
& & \hspace{-.7cm} = - \frac12 u \Bigl[ i\mathcal{B} \!\cdot\! i C_i \!\cdot\!
i\mathcal{B} \Bigr] v \, i\! \mathcal{A}_{ij} \, u \Bigl[ i\mathcal{B}
\!\cdot\! i C_j \!\cdot\! i\mathcal{B} \Bigr] v + i\! \mathcal{A}_{ij} \,
u \Bigl[ i\mathcal{B} \!\cdot\! i C_i i\mathcal{B} i C_j \!\cdot\!
i\mathcal{B} \Bigr] v \qquad \nonumber \\
& & \hspace{0cm} + \frac78 \Biggl(u \Bigl[ i\mathcal{B} \!\cdot\!
i C_i \!\cdot\! i\mathcal{B} \Bigr] v \, i\! \mathcal{A}_{ij} \, u \Bigl[
i\mathcal{B} \!\cdot\! i C_j \!\cdot\! i\mathcal{B} \Bigr] v \Biggr)^{\!2}
\!\! - 3 u \Bigl[ i\mathcal{B} \!\cdot\! i C_i \!\cdot\! i\mathcal{B} \Bigr] v
\, i \! \mathcal{A}_{ij} \nonumber \\
& & \hspace{0cm} \times u \Bigl[ i\mathcal{B} \!\cdot\! i C_j \!\cdot\!
i\mathcal{B} \!\cdot\! i C_k \!\cdot\! i\mathcal{B} \Bigr] v \, i\!
\mathcal{A}_{k\ell} \, u \Bigl[ i\mathcal{B} \!\cdot\! i C_{\ell} \!\cdot\!
i\mathcal{B} \Bigr] v - u \Bigl[ i\mathcal{B} \!\cdot\! i C_i \!\cdot\!
i\mathcal{B} \Bigr] v \nonumber \\
& & \hspace{0cm} \times i\! \mathcal{A}_{ij} \, u \Bigl[ i\mathcal{B}
\!\cdot\! i C_j \!\cdot\!  i\mathcal{B} \Bigr] v \, i\! \mathcal{A}_{k\ell}
\, u \Bigl[ i\mathcal{B} \!\cdot\! i C_k \!\cdot\! i\mathcal{B} \!\cdot\!
i C_{\ell} \!\cdot\! i\mathcal{B} \Bigr] v + u \Bigl[ i\mathcal{B}
\!\cdot\! C_i \!\cdot\! i\mathcal{B} \Bigr] v \, i\! \mathcal{A}_{ij} \,
i\! \mathcal{A}_{k\ell} \nonumber \\
& & \hspace{-.4cm} \times u \Biggl[ i\mathcal{B}
\Biggr(\! i C_j \!\cdot\! i\mathcal{B} \!\cdot\! i C_k \!\cdot\! i\mathcal{B}
\!\cdot\! i C_{\ell} \!+\! i C_k \!\cdot\! i\mathcal{B} \!\cdot\! i C_j
\!\cdot\! i\mathcal{B} \!\cdot\! i C_{\ell} \!+\! i C_k \!\cdot\!
i\mathcal{B} \!\cdot\! i C_{\ell} \!\cdot\! i\mathcal{B} \!\cdot\! i C_j \!
\Biggr) i\mathcal{B} \Biggr] v \nonumber \\
& & \hspace{0cm} + \frac12 i\! \mathcal{A}_{ij} \, u \Bigl[ i\mathcal{B}
\!\cdot\! i C_j \!\cdot\! i\mathcal{B} \!\cdot\! i C_k \!\cdot\!
i\mathcal{B} \Bigr] v \, i\! \mathcal{A}_{k\ell} \, u \Bigl[ i\mathcal{B}
\Bigl( i C_{\ell} \!\cdot\! i\mathcal{B} \!\cdot\! i C_i \!+\! i C_i
\!\cdot\! i\mathcal{B} \!\cdot\! i C_{\ell} \Bigr) i\mathcal{B} \Bigr] v
\nonumber \\
& & \hspace{4cm} + \frac12 \Biggl( i\! \mathcal{A}_{ij} \, u \Bigl[
i\mathcal{B} \!\cdot\! i C_i \!\cdot\! i\mathcal{B} \!\cdot\! i C_j
\!\cdot\! i\mathcal{B} \Bigr] v \Biggr)^{\!2} + O(C^6) \; . \qquad
\label{Jay1}
\end{eqnarray}
Fig.~\ref{fig4} gives a diagrammatic representation of the expansion.

\subsection{Implications for Our Expansion}

Let us compare the new expansion $\mathcal{J} = \mathcal{J}_0 +
\mathcal{J}_1 + \cdots$ with the old one, $\mathcal{J} = J_0 + J_1 +
\cdots$. The first two terms of the new expansion are expressions
(\ref{Jay0}) and (\ref{Jay1}). In contrast, the first two terms of
the old expansion are,
\begin{equation}
J_0 = 1 \qquad , \qquad J_1 = i\!\mathcal{A}_{ij} \, u\Bigl[
i\mathcal{B} \!\cdot\! i C_i \!\cdot\! i\mathcal{B} \!\cdot\! i C_j
i\mathcal{B} \Bigr] v \; .
\end{equation}
It is useful to define the difference between new and old at order
$\ell$,
\begin{equation}
\Delta \! \mathcal{J}_{\ell} \equiv \mathcal{J}_{\ell} - J_{\ell} \;
.
\end{equation}
We can read off $\Delta\! \mathcal{J}_0$ from expression
(\ref{Jay0}),
\begin{eqnarray}
\lefteqn{ \Delta \!\mathcal{J}_0 = \frac12 u \Bigl[ i\mathcal{B}
\!\cdot\! i C_i \!\cdot\! i\mathcal{B} \Bigr] v \,
i\!\mathcal{A}_{ij} \, u \Bigl[ i\mathcal{B} \!\cdot\! i C_j
\!\cdot\! i\mathcal{B} \Bigr] v } \nonumber \\
& & \hspace{.5cm} -\frac38 \Biggl( u \Bigl[ i\mathcal{B} \!\cdot\! i
C_i \!\cdot\! i\mathcal{B} \Bigr] v \, i \! \mathcal{A}_{ij} \, u
\Bigl[ i\mathcal{B} \!\cdot\! i C_j \!\cdot\! i\mathcal{B} \Bigr] v
\Biggr)^2 u \Bigl[ i\mathcal{B} \!\cdot\! i C_i \!\cdot\!
i\mathcal{B} \Bigr] v \, i \! \mathcal{A}_{ij} \nonumber \\
& & \hspace{2.5cm} \times u \Bigl[ i\mathcal{B} \!\cdot\! i C_j
\!\cdot\! i\mathcal{B} \!\cdot\! i C_k \!\cdot\! i\mathcal{B} \Bigr]
v \, i\! \mathcal{A}_{k\ell} \, u \Bigl[ i\mathcal{B} \!\cdot\! i
C_{\ell} \!\cdot\! i\mathcal{B} \Bigr] v + O(C^6) \; . \qquad
\end{eqnarray}
Note that each contribution to $\Delta \!\mathcal{J}_0$ is a ``tree
diagram''; the expansion in powers of the interaction $C$ could also
be viewed as an expansion in numbers of external lines. From
 expression (\ref{Jay1}) we see that $\Delta \! \mathcal{J}_1$
possesses both ``tree'' and ``one loop'' contributions,
\begin{eqnarray}
\lefteqn{ \Delta \!\mathcal{J}_1^0 = -\frac12 u \Bigl[ i\mathcal{B}
\!\cdot\! i C_i \!\cdot\! i\mathcal{B} \Bigr] v \,
i\!\mathcal{A}_{ij} \, u \Bigl[ i\mathcal{B} \!\cdot\! i C_j
\!\cdot\! i\mathcal{B} \Bigr] v } \nonumber \\
& & \hspace{.5cm} +\frac78 \Biggl( u \Bigl[ i\mathcal{B} \!\cdot\! i
C_i \!\cdot\! i\mathcal{B} \Bigr] v \, i \! \mathcal{A}_{ij} \, u
\Bigl[ i\mathcal{B} \!\cdot\! i C_j \!\cdot\! i\mathcal{B} \Bigr] v
\Biggr)^2 - 3 u \Bigl[ i\mathcal{B} \!\cdot\! i C_i \!\cdot\!
i\mathcal{B} \Bigr] v \, i \! \mathcal{A}_{ij} \nonumber \\
& & \hspace{2.5cm} \times u \Bigl[ i\mathcal{B} \!\cdot\! i C_j
\!\cdot\! i\mathcal{B} \!\cdot\! i C_k \!\cdot\! i\mathcal{B} \Bigr]
v \, i\! \mathcal{A}_{k\ell} \, u \Bigl[ i\mathcal{B} \!\cdot\! i
C_{\ell} \!\cdot\! i\mathcal{B} \Bigr] v + O(C^6) \; , \qquad \\
\lefteqn{ \Delta \!\mathcal{J}_1^1 = - u \Bigl[ i\mathcal{B}
\!\cdot\! i C_i \!\cdot\! i\mathcal{B} \Bigr] v i\! \mathcal{A}_{ij}
\, u \Bigl[ i\mathcal{B} \!\cdot\! i C_j \!\cdot\! i\mathcal{B}
\Bigr] v \, i\! \mathcal{A}_{k\ell} \, u \Bigl[ i\mathcal{B}
\!\cdot\! i C_k \!\cdot\! i\mathcal{B} \!\cdot\! i C_{\ell}
\!\cdot\! i\mathcal{B} \Bigr] v } \nonumber \\
& & \hspace{-.45cm} + u \Bigl[ i\mathcal{B} \!\cdot\! C_i \!\cdot\!
i\mathcal{B} \Bigr] v \, i\! \mathcal{A}_{ij} \, i\!
\mathcal{A}_{k\ell} \, u \Biggl[ i\mathcal{B} \Biggr(\! i C_j
\!\cdot\! i\mathcal{B} \!\cdot\! i C_k \!\cdot\! i\mathcal{B}
\!\cdot\! i C_{\ell} \!+\! i C_k \!\cdot\! i\mathcal{B} \!\cdot\! i
C_j \!\cdot\! i\mathcal{B} \!\cdot\! i C_{\ell} \nonumber \\
& & \hspace{-.45cm} + i C_k \!\cdot\! i\mathcal{B} \!\cdot\! i
C_{\ell} \!\cdot\! i\mathcal{B} \!\cdot\! i C_j \! \Biggr)
i\mathcal{B} \Biggr] v + \frac12 i\! \mathcal{A}_{ij} \, u \Bigl[
i\mathcal{B} \!\cdot\! i C_j \!\cdot\! i\mathcal{B} \!\cdot\! i C_k
\!\cdot\! i\mathcal{B} \Bigr] v \, i\! \mathcal{A}_{k\ell} \, u
\Bigl[ i\mathcal{B} \Bigl(
i C_{\ell} \nonumber \\
& & \hspace{-.45cm} \times i\mathcal{B} \!\cdot\! i C_i + i C_i
\!\cdot\! i\mathcal{B} \!\cdot\! i C_{\ell} \Bigr) i\mathcal{B}
\Bigr] v \!+\! \frac12 \Biggl( i\! \mathcal{A}_{ij} \, u \Bigl[
i\mathcal{B} \!\cdot\! i C_i \!\cdot\! i\mathcal{B} \!\cdot\! i C_j
\!\cdot\! i\mathcal{B} \Bigr] v \Biggr)^{\!2} \!\! + O(C^6) \; .
\qquad
\end{eqnarray}

Note that the order $C^2$ contributions to $\Delta \!\mathcal{J}_0$
and $\Delta \!\mathcal{J}_1^0$ cancel. They do not cancel at order
$C^4$, but the residual $C^4$ terms are canceled by $\Delta \!
\mathcal{J}_2^0$. So the new expansion does not move the old diagrams
from one order to another, rather it adds a new class of diagrams
--- with more external lines --- to one order and subtracts them from
higher orders. What we have learned can be summarized by the following
observations:
\begin{itemize}
\item{Each term $\mathcal{J}_{\ell}$ in the new expansion can be
written as an infinite series which begins at order $C^{2\ell}$;}
\item{The lowest order term in this series expansion of each
$\mathcal{J}_{\ell}$ is the old $\ell$-loop term $J_{\ell}$;}
\item{The new terms in $\mathcal{J}_{\ell}$ involve diagrams with
up to and including $\ell$ loops;}
\item{The new terms at each order have more powers of the
external state factors $u^a$ and $v^b$; and}
\item{The sum of all the new terms is zero.}
\end{itemize}

Although there is no guarantee that the new expansion is free of
ultraviolet divergences, it does possess a number of desirable
features. Note first that ultraviolet divergences are no worse in
the new expansion than the old one because none of the new diagrams
in $\mathcal{J}_{\ell}$ have more than $\ell$ loops. Our simple
model makes no distinction between photons and gravitons but the
quantum field theoretic expansion will of course involve both
particles. The new diagrams --- with more powers of the coupling
constants $e^2$ and $G$, but no more loops --- are one way gravity
at order $\ell$ in the new expansion can know about an order $\ell$
divergence from the gauge sector. So the new expansion breaks the
conundrum of conventional perturbation theory that the gravitational
response to a problem at one order cannot come until the next order.
Note finally, that the infinite sums of new diagrams at any fixed
order of the new expansion allow the possibility of getting
nonanalytic dependence upon $G$ and $e^2$.

\section{The 0th Order Term}

The purpose of this section is the evaluate the 0th order term in
the new expansion of expression (\ref{finalform}) for the scalar
mass. Recall that this means solving the problem of a first
quantized Klein-Gordon scalar which forms a bound state in the
gravitational and electrostatic potentials sourced by its own
probability current. We begin by specializing the Lagrangians, the
field equations and the normalization condition to the bound state
ansatz (\ref{ansatz1}-\ref{ansatz3}). We then note that one of the
equations can be solved exactly for the negative energy gravitational
degree of freedom. Substituting this solution into the action gives a
functional of the remaining fields whose minimization yields the
remaining field equations. This functional serves as the basis for a
variational formulation of the problem. The section closes with a
numerical solution.

\subsection{Field Equations}

If a bound state forms we can simplify the fields to take the form
(\ref{ansatz1}-\ref{ansatz3}),
\begin{equation}
\phi = e^{-i E t} F(r) \qquad , \qquad A_{\mu} = -\Phi(r)
\delta^0_{\mu} \; , \label{FandPhi}
\end{equation}
\begin{equation}
g_{00} = -B(r) \qquad , \qquad g_{0i} = 0 \qquad , \qquad g_{ij} =
\pi_{ij} + \widehat{r}_i \widehat{r}_j A(r) \; . \label{AandB}
\end{equation}
Here $\widehat{r}_i \equiv x_i/r$ is the radial unit vector and
$\pi_{ij} \equiv \delta_{ij} - \widehat{r}_i \widehat{r}_j$ is the
transverse projection operator. The nonvanishing components of the
affine connection are,
\begin{equation}
\Gamma^0_{~0i} = \widehat{r}_i \Bigl(\frac{B'}{2B}\Bigr) \quad ,
\quad \Gamma^i_{~00} = \widehat{r}_i \Bigl(\frac{B'}{2 A}\Bigr)
\quad , \quad \Gamma^{i}_{~jk} = \widehat{r}_i \widehat{r}_j
\widehat{r}_k \frac{A'}{2 A} \!+\! \widehat{r}_i \pi_{jk} \Bigl(
\frac{A \!-\! 1}{r A} \Bigr) \; .
\end{equation}
The nonzero components of the Riemann tensor are,
\begin{eqnarray}
R^0_{~i0j} & = & \widehat{r}_i \widehat{r}_j \Bigl[ -\frac{B''}{2 B}
\!+\! \frac{ {B'}^2}{4 B^2} \!+\! \frac{A' B'}{4 A B} \Bigr] -
\pi_{ij} \Bigl( \frac{B'}{2 r A B} \Bigr) \; , \qquad \\
R^i_{~jk\ell} & = & \Bigl[ \pi_{ik} \widehat{r}_j \widehat{r}_{\ell}
\!-\! \pi_{i\ell} \widehat{r}_j \widehat{r}_k \Bigr]
\Bigl(\frac{A'}{2 r A} \Bigr) + \Bigl[ \pi_{i\ell} \widehat{r}_i
\widehat{r}_k \!-\! \pi_{jk} \widehat{r}_i \widehat{r}_{\ell} \Bigr]
\Bigl(\frac{A'}{2 r A^2} \Bigr) \qquad \nonumber \\
& & \hspace{6cm} + \Bigl[ \pi_{ik} \pi_{j\ell} \!-\! \pi_{i\ell}
\pi_{jk} \Bigr] \Bigl( \frac{A \!-\! 1}{r^2 A}\Bigr) \; . \qquad
\end{eqnarray}
Contraction gives the Ricci tensor $R_{\mu\nu} =
R^{\rho}_{~\mu\rho\nu}$,
\begin{eqnarray}
R_{00} & = & \frac{B''}{2A} \!-\! \frac{{B'}^2}{4 A B} \!-\!
\frac{A' B'}{4 A^2} \!+\! \frac{A'}{r A} \; , \qquad \\
R_{ij} & = & \widehat{r}_i \widehat{r}_j \Bigl[ -\frac{B''}{2 B}
\!+\! \frac{ {B'}^2}{4 B^2} \!+\! \frac{A' B'}{4 A B} \!+\!
\frac{A'}{r A} \Bigr] \nonumber \\
& & \hspace{4.5cm} + \pi_{ij} \Bigl[ -\frac{B'}{2 r A B} \!+\!
\frac{A'}{2 r A^2} + \Bigl( \frac{A \!-\! 1}{r^2 A}\Bigr) \Bigr] \;
. \qquad
\end{eqnarray}
And contracting that gives the Ricci scalar $R = g^{\mu\nu}
R_{\mu\nu}$,
\begin{equation}
R = -\frac{B''}{A B} \!+\! \frac{ {B'}^2}{2 A B^2} \!+\! \frac{A'
B'}{2 A^2 B} \!+\! \frac{2 A'}{r A^2} \!-\! \frac{2 B'}{r A B} \!+\!
2 \Bigl( \frac{A \!-\! 1}{r^2 A} \Bigr) \; .
\end{equation}

In specializing the various Lagrangians (\ref{LGR}-\ref{LSC}) to our
ansatz (\ref{FandPhi}-\ref{AandB}) it is best to use spherical
coordinates and integrate the angles so that the effective measure
factor is,
\begin{equation}
\sqrt{-g} = 4\pi r^2 \sqrt{A(r) B(r)} \; .
\end{equation}
ADM long ago worked out the surface term one must add to the Hilbert
Lagrangian (\ref{LGR}) for asymptotically flat field configurations
\cite{ADM2}. For our static, spherically symmetric geometry the
result is,
\begin{equation}
\mathcal{L}_{\rm ADM} \longrightarrow \frac1{16 \pi G} \Biggl[ R
\sqrt{-g} - \frac{\partial}{\partial r} \Bigl( \frac{B' \sqrt{-g}}{A
B} \Bigr) \Biggr] = \frac1{8 \pi G} \Biggl[ \frac{A'}{r A^2} \!+\!
\Bigl(\frac{A \!-\! 1}{r^2 A}\Bigr) \Biggr] \sqrt{-g} \; .
\end{equation}
The other two Lagrangians (\ref{LEM}-\ref{LSC}) require no surface
terms,
\begin{eqnarray}
\mathcal{L}_{\rm EM} & \longrightarrow & \frac{\epsilon_0 {\Phi'}^2}{2 A B}
\sqrt{-g} \; , \qquad \\
\mathcal{L}_{\rm SC} & \longrightarrow & \frac1{B} (E \!+\! e
\Phi)^2 F^2 \sqrt{-g} -\frac1{A} {F'}^2 \sqrt{-g} - M_0^2 F^2
\sqrt{-g} \; . \qquad
\end{eqnarray}
Hence the action is,
\begin{eqnarray}
\lefteqn{S[F,\Phi,B,A] = (t_{+} \!-\! t_{-}) \int_0^{\infty} \!\! dr
\, \sqrt{-g} \, \Biggl\{ \frac1{8 \pi G} \Biggl[ \frac{A'}{r A^2}
\!+\! \Bigl(\frac{A \!-\! 1}{r^2 A}\Bigr) \Biggr] } \nonumber \\
& & \hspace{4.5cm} + \frac{\epsilon_0 {\Phi'}^2}{2 A B} + \frac{(E \!+\! e
\Phi)^2 F^2}{B} - \frac{{F'}^2}{A} - M_0^2 F^2 \Biggr\} . \qquad
\label{Strunc}
\end{eqnarray}

It is well known that the operation of making an ansatz for the
solution does not commute with varying the action to obtain the
field equations. The correct procedure is to vary first and then
simplify. However, all of the equations one gets by simplifying
first and then varying are correct, and it turns out that the only
ones we miss are the trivial relations implied by the Bianchi
identity \cite{Palais}. Except for the overall factor of $\Delta t
\equiv (t_{+} - t_{-})$, the variations with respect to various
fields are,
\begin{eqnarray}
\frac1{\Delta t} \frac{\delta S}{\delta F} & = &
\frac{\partial}{\partial r} \Biggl[ \frac{2 F' \sqrt{-g}}{A} \Biggr]
+ \frac{2 (E \!+\! e \Phi)^2 F \sqrt{-g}}{B} - 2 M_0^2 F \sqrt{-g}
\; , \qquad \label{FEQN} \\
\frac1{\Delta t} \frac{\delta S}{\delta \Phi} & = &
\frac{\partial}{\partial r} \Biggl[ \frac{\epsilon_0 \Phi' \sqrt{-g}}{A B}
\Biggr] + \frac{2 e (E \!+\! e \Phi) F^2 \sqrt{-g}}{B} \; , \qquad
\label{PHIEQN} \\
\frac1{\Delta t} \frac{\delta S}{\delta B} & = & \Biggl\{
\frac1{8 \pi G} \Biggl[ \frac{A'}{r A^2} \!+\! \Bigl( \frac{A \!-\!
1}{r^2 A} \Bigr) \Biggr] - \frac{\epsilon_0 {\Phi'}^2}{2 A B} \nonumber \\
& & \hspace{3.5cm} - \frac{(E \!+\! e \Phi)^2 F^2}{B} - \frac{{F'}^2}{A}
- M_0^2 F^2 \Biggr\} \frac{\sqrt{-g}}{2 B} \; , \qquad \label{BEQN} \\
\frac1{\Delta t} \frac{\delta S}{\delta A} & = & \Biggl\{
\frac1{8 \pi G} \Biggl[-\frac{B'}{r A B} \!+\! \Bigl( \frac{A \!-\!
1}{r^2 A} \Bigr) \Biggr] - \frac{\epsilon_0 {\Phi'}^2}{2 A B} \nonumber \\
& & \hspace{3.5cm} + \frac{(E \!+\! e \Phi)^2 F^2}{B} + \frac{{F'}^2}{A}
- M_0^2 F^2 \Biggr\} \frac{\sqrt{-g}}{2 A} \; . \qquad \label{AEQN}
\end{eqnarray}

Our goal is to find $F(r)$, $\Phi(r)$, $B(r)$ and $A(r)$ so as to
make each of the variations (\ref{FEQN}-\ref{AEQN}) vanish. However,
those equations have solutions for any constant $E$. As always, it
is normalizability which puts the ``quantum'' in quantum mechanics.
For our problem the normalization condition is,
\begin{equation}
2 \int_0^{\infty} \!\! dr \sqrt{-g} \, \frac{(E \!+\! e \Phi) F^2}{B}
= 1 \; . \label{norm}
\end{equation}
Any field configurations $F_0(r)$, $\Phi_0(r)$, $B_0(r)$ and
$A_0(r)$ which make all the variations (\ref{FEQN}-\ref{AEQN})
vanish {\it and} also obey (\ref{norm}) will only do so for very
special values of $E$. Note that the field equations and
normalizability requires $A(r)$ to approach one at infinity, and
$F(r)$ to approach zero. The asymptotic values of $\Phi(r)$ and
$B(r)$ are both gauge choices. By choosing the $U(1)$ gauge
parameter to be $-t \times \Phi_{\infty}$ we can make $\Phi(r)$
vanish at infinity. By changing time to $t/k$ we induce the
rescalings,
\begin{eqnarray}
B(r) & \longrightarrow & k^2 \times B(r) \; , \\
\Phi(r) & \longrightarrow & k \times \Phi(r) \; , \\
E & \longrightarrow & k \times E \; .
\end{eqnarray}
We shall always use this freedom to make $B(r)$ approach one at
infinity.

If we can find a normalized solution $F_0(r)$, $\Phi_0(r)$, $B_0(r)$
and $A_0(r)$ then our zeroth order result for the mass is,
\begin{equation}
M_{\rm 0th} = E - \frac1{\Delta t} S[F_0,\Phi_0,B_0,A_0] \; .
\label{answer}
\end{equation}
The first term on the right hand side of (\ref{answer}) is from the
the two scalar fields,
\begin{equation}
\phi(t_{+},\vec{x}) \phi^*(t_{-},\vec{0}) \longrightarrow F_0(r)
F_0(0) e^{-i E \Delta t} \; .
\end{equation}
The final term in (\ref{answer}) represents the gravitational and
electromagnetic contribution to the mass. Note that the scalar
action vanishes for solutions because the scalar Lagrangian is a
surface term which goes to zero,
\begin{equation}
\mathcal{L}_{\rm SC} \longrightarrow -\frac{\partial}{\partial r}
\Biggl[ \frac{F_0 F_0' \sqrt{-g_0}}{A_0} \Biggr] \; .
\end{equation}

\subsection{A Variational Formalism}

Solving differential equations is tough, and we are not able to
find exact solutions for all four of the fields. For many bound state
problems in quantum mechanics the absence of exact solutions is not
crippling because variational techniques allow one to derive strong
bounds on the ground state energy. Such a technique would be simple
to formulate for our Klein-Gordon scalar if only the electromagnetic
and gravitational potentials were fixed. However, the fact that these
potentials are sourced by the Klein-Gordon wave function itself
endows this problem with a slippery, nonlinear character. The presence
of gravitational interactions is especially problematic because some
of the constrained degrees of freedom in gravity possess negative
energy. Instability is only avoided by constraining these degrees of
freedom to obey their field equations; attempting to minimize the
action with respect to these degrees of freedom would carry one away
from the actual solution.

There are good reasons for suspecting that the field $B(r)$ is the
only negative energy degree of freedom. In the normal ADM formalism
$B = N^2$ would be the square of the lapse field, and it could be
specified arbitrarily as a choice of gauge. However, $B$ is a dynamical
degree of freedom in this problem. The structure of our Lagrangian is
similar to the usual formalism for describing cosmological
perturbations during primordial inflation \cite{KOW}. In that setting,
as for us, the Lagrangian can be written as the sum of a ``kinetic''
part $K$ and a ``potential'' part $P$,
\begin{equation}
\mathcal{L} = \Biggl[ \frac{K}{B} + P\Biggr] \sqrt{-g} \; .
\label{Lagrangian}
\end{equation}
For us the kinetic and potential parts are,
\begin{eqnarray}
K & = & \frac{\epsilon_0 {\Phi'}^2}{2 A} + (E \!+\! e \Phi)^2 F^2 \; ,
\qquad \label{kinetic} \\
P & = & \frac1{8 \pi G} \Biggl[ \frac{A'}{r A^2} \!+\! \Bigl( \frac{A
\!-\! 1}{r^2 A} \Bigr) \Biggr] - \frac{ {F'}^2}{A} - M_0^2 F^2 \; .
\qquad \label{potential}
\end{eqnarray}
The field equation for $B$ is algebraic and has a trivial solution,
\begin{equation}
\frac1{\Delta t} \frac{\delta S}{\delta B} = \Biggl[ -\frac{K}{B} + P
\Biggr] \frac{\sqrt{-g}}{2 B} = 0 \qquad \Longrightarrow \qquad
B = \frac{K}{P} \; . \label{Bsolution}
\end{equation}

Substituting (\ref{Bsolution}) into (\ref{Lagrangian}) allows
us to express the action in terms of just $F(r)$, $\Phi(r)$ and
$A(r)$,
\begin{equation}
S[F,\Phi,A] = 8 \pi \Delta t \int_0^{\infty} \!\! dr \, r^2
\sqrt{A K P} \; . \label{newS}
\end{equation}
It is simple to show that varying (\ref{newS}) gives equations
(\ref{FEQN}-\ref{PHIEQN}) and (\ref{AEQN}). Because (\ref{newS})
is positive semi-definite, the problem of extremizing it is likely
to be the same as that of minimizing it. The corresponding normalization
condition is,
\begin{equation}
8 \pi \int_0^{\infty} \!\! dr \, r^2 \sqrt{\frac{A K}{P}} \,
(E \!+\! e \Phi) F^2 = 1 \; .
\end{equation}
And our zeroth order result for the scalar mass becomes,
\begin{equation}
M_{\rm 0th} = E - 8 \pi \int_0^{\infty} \!\! dr \, r^2 \sqrt{ A_0
K_0 P_0} \; .
\end{equation}

We illustrate the method with simple trial functions for $F(r)$,
$\Phi(r)$ and $A(r)$. We cannot make $F(r)$ a spherical shell like
ADM, or even a hard sphere, because the factors of ${F'}^2$ become
ill-defined if $F(r)$ has a discontinuity. The next best thing is
to assume the scalar drops linearly to zero within some distance $R$,
\begin{equation}
F(r) = a (R \!-\! r) \theta(R \!-\! r) \; . \label{Fansatz}
\end{equation}
Comparably simple forms for the potentials are,
\begin{eqnarray}
\Phi(r) & = & -\frac{e \theta(R \!-\! r)}{4 \pi \epsilon_0 R}
- \frac{e \theta(r \!-\! R)}{4 \pi \epsilon_0 r} \; , \qquad
\label{Phiansatz} \\
\frac1{A(r)} & = & \Bigl[1 \!-\! b r^2\Bigr] \theta(R \!-\! r)
+ \Bigl[1 \!-\! \frac{c}r \!+\! \frac{d}{r^2}\Bigr]
\theta(r \!-\! R) \; . \label{Aansatz} \qquad
\end{eqnarray}
We can make $A(r)$ continuous by choosing,
\begin{equation}
b = \frac{c}{R^3} - \frac{d}{R^4} \; .
\end{equation}
The corresponding forms for the kinetic and potential terms are,
\begin{eqnarray}
K(r) & = & \Bigl(E \!-\! \frac{\alpha}{R}\Bigr)^2 a^2 (R \!-\! r)^2
\theta(R \!-\! r) + \frac{\alpha}{8\pi r^4} \Bigl(1 \!-\! \frac{c}{r}
\!+\! \frac{d}{r^2} \Bigr) \theta(r \!-\! R) \; , \label{Kansatz}
\qquad \\
P(r) & = & \Biggl[ \frac{3 b}{8\pi G} - a^2 \Bigl[1 \!-\! b r^2
\!+\! M_0^2 (R \!-\! r)^2\Bigr] \Biggr] \theta(R \!-\! r) +
\frac{d \theta(r \!-\! R)}{8\pi G r^4} \; . \qquad \label{Pansatz}
\end{eqnarray}
Hence the potential $B(r)$ is,
\begin{equation}
B(r) = \frac{(E \!-\! \frac{\alpha}{R})^2 a^2 (R \!-\! r)^2
\theta(R \!-\! r) }{ \frac{3 b}{8\pi G} - a^2 [1 \!-\! b r^2
\!+\! M_0^2 (R \!-\! r)^2] } + \frac{ \frac{\alpha}{8\pi r^4}
(1 \!-\! \frac{c}{r} \!+\! \frac{d}{r^2} ) \theta(r \!-\! R) }{
\frac{d}{8\pi G r^4} } \; .
\end{equation}
Enforcing that $B(r)$ goes to one at infinity determines the
coefficient $d$,
\begin{equation}
d = \alpha G \; ,
\end{equation}
where $\alpha \equiv e^2/4\pi\epsilon_0 \approx 1/137$ is the fine
structure constant. Enforcing continuity at $r=R$ --- which means $B(R)
= 0$ --- requires the choice,
\begin{equation}
c = R + \frac{\alpha G}{R} \qquad \Longrightarrow \qquad
b = \frac1{R^2} \; .
\end{equation}

At this stage the free parameters in our trial solution are
$R$, $a$ and the energy $E$. The next step is to enforce
normalizability, which requires,
\begin{eqnarray}
\frac1{8\pi} & = & \int_0^{R} \!\! \frac{dr \, r^2}{\sqrt{1 \!-\!
b r^2}} \frac{ (E \!-\! \frac{\alpha}{R})^2 a^3 (R \!-\! r)^3}{
\sqrt{ \frac{3b}{8\pi G} \!-\! a^2 [ 1 \!-\! b r^2 \!+\! M_0^2
(R \!-\! r)^2] } } \; , \qquad \\
& = & (R E \!-\! \alpha)^2 R^4 a^2 \int_0^1 \!\! \frac{dx \, x^2}{
\sqrt{1 \!-\! x^2}} \frac{ (1 \!-\! x)^3}{ \sqrt{
\frac{3}{8\pi G R^2 a^2} \!-\! [ 1 \!-\! x^2 \!+\! R^2 M_0^2
(1 \!-\! x)^2 ]} } \; , \qquad \\
& \equiv & (R E \!-\! \alpha)^2 R^4 a^2 \times
I\Bigl( GR^2 a^2, R^2 M_0^2) \; . \qquad \label{newIdef}
\end{eqnarray}
The function $I(x,y)$ in equation (\ref{newIdef}) can be expressed
in terms of elliptic integrals but we may as well treat it as an
elementary function and use it to express the energy,
\begin{equation}
E = \Biggl[\frac1{\sqrt{8\pi R^4 a^2 I(G R^2 a^2, R^2 M_0^2)}}
+\alpha \Biggr] \frac1{R} \; .
\end{equation}

We can now regard the two free parameters as,
\begin{equation}
A \equiv \frac{R a}{M_0} \qquad , \qquad \mu \equiv R M_0 \; .
\end{equation}
The field action is,
\begin{equation}
\frac{S}{\Delta t} = \Biggl[ \frac{ \sqrt{8\pi} \, A J(G M_0^2
A^2,\mu^2)}{\sqrt{ I(G M_0^2 A^2,\mu^2)} } + \frac{\alpha}{\mu}
\Biggr] M_0 \; , \label{fieldS}
\end{equation}
where the new integral is,
\begin{equation}
J(G M_0^2 A^2,\mu^2) \equiv \int_0^1 \! \frac{dx \, x^2 (1 \!-\!
x)}{ \sqrt{1 \!-\! x^2}} \sqrt{ \frac{3}{8\pi G M_0^2 A^2} \!-\! [1
\!-\! x^2 \!+\! \mu^2 (1 \!-\! x)^2 ] } \; . \label{Jdef}
\end{equation}
We determine the free parameters $A$ and $\mu$ by minimizing
$S/\Delta t$.

Expressions (\ref{newIdef}), (\ref{fieldS}) and (\ref{Jdef}) seem
very complicated. However, note that because $I(G M_0^2 A^2,\mu^2)$
is an increasing function of $A$ and $\mu$, and $J(G M_0^2 A^2,
\mu^2)$ is a decreasing function of $A$ and $\mu$, $S/\Delta t$ is a
decreasing function of $\mu$ at fixed $A$. Hence $S/\Delta t$ is
minimized, at fixed $A$, by choosing $\mu$ to be the maximum value
for which the two integrals remain real,
\begin{equation}
\mu_{\rm max} = \sqrt{ \frac{3}{8\pi G M_0^2 A^2} \!-\! 1} \qquad
\Longrightarrow \qquad 0 < A < A_{\rm max} = \sqrt{\frac{3}{8\pi G
M_0^2}} \; .
\end{equation}
At $\mu = \mu_{\rm max}$ the two integrals become,
\begin{eqnarray}
i(\mu_{\rm max}) & \equiv & I(G M_0^2 A^2,\mu_{\rm max}^2) = \int_0^1 \!\! 
\frac{dx \, x^2}{\sqrt{1 \!-\! x^2}} \frac{ (1 \!-\! x)^3}{ \sqrt{ \mu_{\rm
max}^2 [1 \!-\! (1 \!-\! x)^2] \!+\! x^2} } \; , \qquad \label{Imax} \\
j(\mu_{\rm max}) & \equiv & J(G M_0^2 A^2, \mu_{\rm max}^2) = \int_0^1 \! 
\frac{dx \, x^2 (1 \!-\! x)}{\sqrt{1 \!-\! x^2}} \sqrt{ \mu_{\rm max}^2 
[1 \!-\! (1 \!-\! x)^2] \!+\! x^2} \; . \qquad \label{Jmax}
\end{eqnarray}
And the field action (\ref{fieldS}) takes the form,
\begin{equation}
\frac{S}{\Delta t} = \Biggl[ \sqrt{ \frac{3}{G M_0^2} } \,
\frac{ j(\mu_{\rm max}) }{ \sqrt{(1 \!+\! \mu_{\rm max}^2) 
i(\mu_{\rm max}) } } + \frac{\alpha}{\mu_{\rm max}} \Biggr] M_0 \; .
\label{finalS}
\end{equation}

\begin{figure}
\includegraphics[width=8cm,height=8cm]{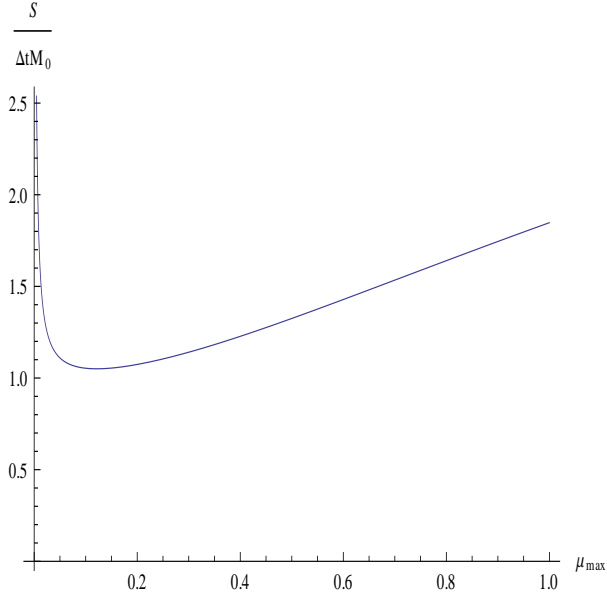}
\caption{Plot of $\frac{S}{M_0\Delta t}$ versus $\mu_{\rm max}$ from equation
(\ref{finalS}) for $G M_0^2 = 0.36$. The minimum seems to be at about 
$\mu_{\rm max} = 0.12$.}
\label{Svsmu}
\end{figure}

At this stage the problem is numerical. Fig.~\ref{Svsmu} shows 
$S/M_0 \Delta t$ as a function of $\mu_{\rm max}$ for $G M_0^2 =
0.36$. The minimum seems to be at about $\mu_{\rm max} = 0.12$,
which corresponds to,
\begin{eqnarray}
R & = & \frac{\mu_{\rm max}}{\sqrt{G M_0^2}} \times \sqrt{G} 
\approx 0.20 \times \sqrt{G} \; , \qquad \label{res1} \\
a & = & \sqrt{\frac{3 G M_0^2}{8\pi} } \frac1{ \mu_{\rm max}
\sqrt{1 \!+\! \mu_{\rm max}^2 }} \times \frac1{G} 
\approx 1.7 \times \frac1{G} \; , \qquad \label{res2} \\
E & = & \Biggl[ \frac{G M_0^2}{\mu_{\rm max}^2} \sqrt{ \frac{1 \!+\! 
\mu_{\rm max}^2 }{ 3 i(\mu_{\rm max}) }} \!+\!
\frac{\alpha \sqrt{G M_0^2}}{\mu_{\rm max}} \Biggr] \frac1{\sqrt{G}}
\approx 62 \times \frac1{\sqrt{G}} \; , \label{res3} \\
M & = & E - \frac{S}{\Delta t} \approx 62 \times \frac1{\sqrt{G}} 
\; . \qquad \label{res4}
\end{eqnarray}
We will see that these results are not very accurate.

\subsection{Numerical Results}

It is desirable to check any variational ansatz against a direct,
numerical solution to the problem. Of course computers can only
solve for dimensionless quantities, so it is first necessary to
express everything in geometrodynamical units, using $G$ to 
absorb each quantity's natural units,
\begin{equation}
r = \sqrt{G} \, \widetilde{r} \;\; , \;\; 
M_0 = \frac{\widetilde{M}}{\sqrt{G}} \;\; , \;\;
E = \frac{\widetilde{E}}{\sqrt{G}} \;\; , \;\;
M = \frac{\widetilde{M}}{\sqrt{G}} \;\; ,
\end{equation}
\begin{equation}
F(r) = \frac{\widetilde{F}(\widetilde{r})}{\sqrt{G}} \;\; , \;\;
\Phi(r) = \frac{[\widetilde{\Phi}(\widetilde{r}) \!-\! \widetilde{E}]}{e 
\sqrt{G}} \;\; , \;\;
B(r) = \widetilde{B}(\widetilde{r}) \;\; , \;\;
A(r) = \widetilde{A}(\widetilde{r}) \;\; , 
\end{equation}
\begin{equation}
\sqrt{-g(r)} = G \sqrt{-\widetilde{g}(\widetilde{r})} \;\; , \;\;
K(r) = \frac{\widetilde{K}(\widetilde{r})}{G^2} \;\; , \;\;
P(r) = \frac{\widetilde{P}(\widetilde{r})}{G^2} \;\; .
\end{equation}
Note that we have absorbed the energy into the electrostatic potential.
In all cases we employ a tilde to denote the dimensionless quantity.
Geometrodynamic fields such as $\widetilde{F}$ are considered to be 
functions of the geometrodynamic radius $\widetilde{r}$. A prime on
such a field indicates differentiation with respect to $\widetilde{r}$,
so we have,
\begin{equation}
F' = \frac{\widetilde{F}'}{G} \qquad , \qquad \Phi' = 
\frac{ \widetilde{\Phi}'}{e G} \; .
\end{equation}

\begin{figure}
\includegraphics[width=6.5cm,height=6.5cm]{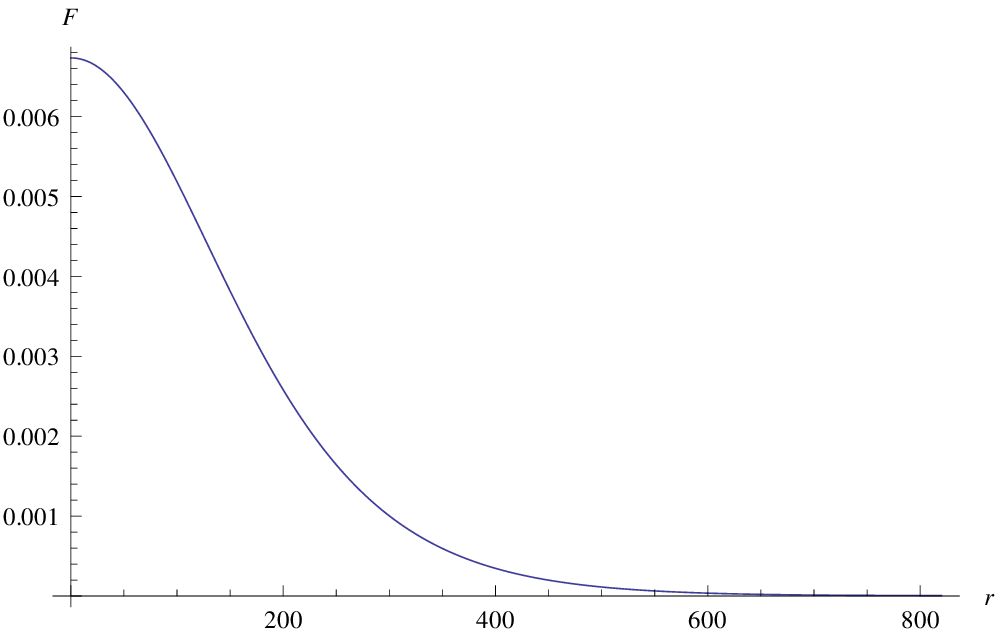}
\hspace{.5cm}
\includegraphics[width=6.5cm,height=6.5cm]{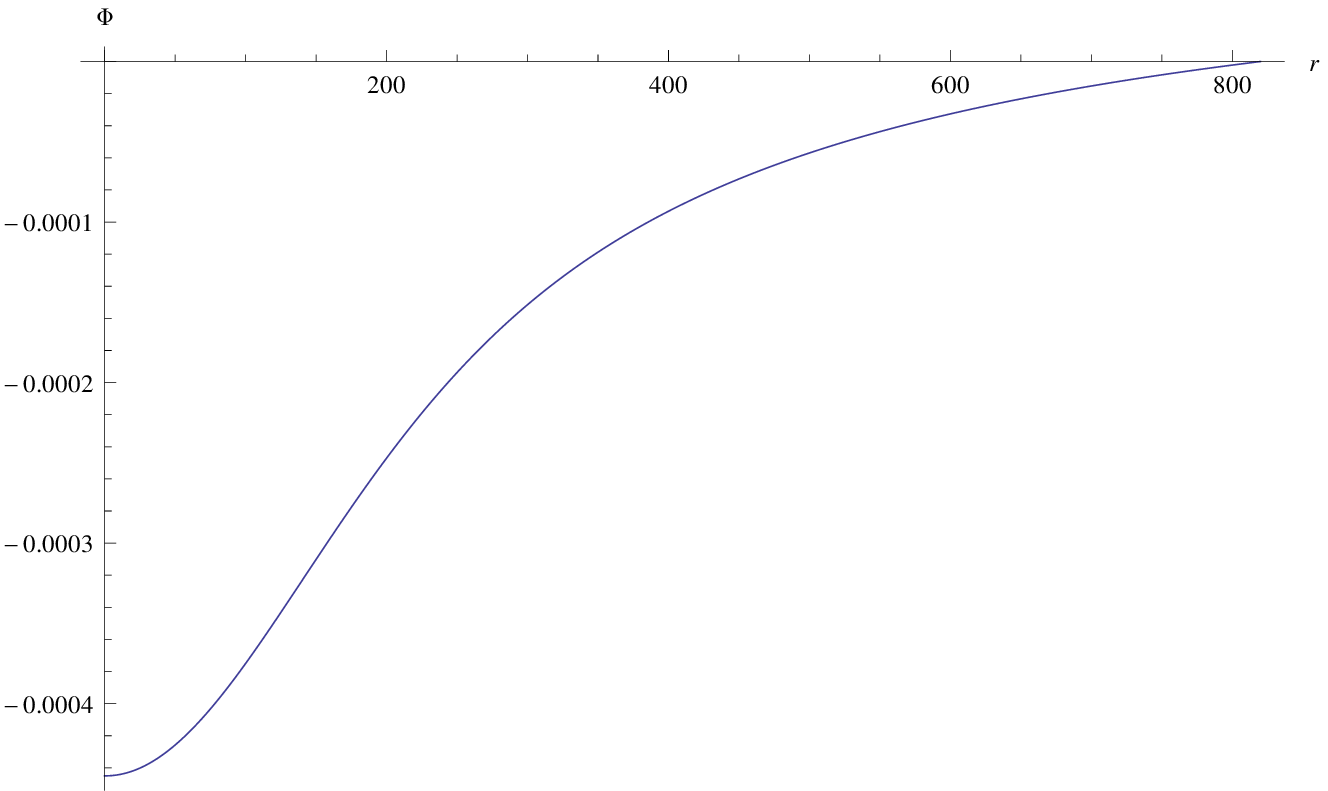}
\caption{Plots of the scalar amplitude $F(r)$ (in units of $M_{\rm Pl}
= 1/\sqrt{G}$) and the electrostatic potential $\Phi(r)$ (in units of 
$M_{\rm Pl}/e$) as functions of $r$ (in units of $1/M_{\rm Pl}$). 
These figures were generated for bare mass $M_0 = 0.60 \, M_{\rm Pl}$.}
\label{FandPhiplots}
\end{figure}

In these units the four field equations (\ref{FEQN}-\ref{AEQN}) take
the form,
\begin{eqnarray}
& & \hspace{-.5cm} \frac{\partial}{\partial \widetilde{r}} \Biggl[ \frac{ 
\widetilde{F}' \sqrt{-\widetilde{g}} }{\widetilde{A}} \Biggr]
+ \frac{\widetilde{\Phi}^2 \widetilde{F}
\sqrt{-\widetilde{g}} }{\widetilde{B}} - \widetilde{M}_0^2 
\widetilde{F} \sqrt{-\widetilde{g}} = 0 \; , \qquad \label{FTEQN} \\
& & \hspace{-.5cm} \frac{\partial}{\partial \widetilde{r}} \Biggl[ \frac{ 
\widetilde{\Phi}' \sqrt{-\widetilde{g}} }{\widetilde{A} \widetilde{B}} \Biggr]
+ \frac{8\pi \alpha \widetilde{\Phi} \widetilde{F}^2
\sqrt{-\widetilde{g}} }{\widetilde{B}} = 0 \; , \qquad \label{PTEQN} \\
& & \hspace{-.5cm} \frac1{8 \pi} \Biggl[ \frac{ \widetilde{A}' }{ \widetilde{r}
\widetilde{A}^2 } \!+\! \Bigl(\frac{ \widetilde{A} \!-\! 1}{
\widetilde{r}^2 \widetilde{A} } \Bigr) \Biggr] - 
\frac{ \widetilde{\Phi}^{\prime 2} }{8\pi \alpha \widetilde{A} \widetilde{B} }
- \frac{ \widetilde{\Phi}^2 \widetilde{F}^2}{ \widetilde{B}} - 
\frac{ \widetilde{F}^{\prime 2} }{\widetilde{A}}
- \widetilde{M}_0^2 \widetilde{F}^2 = 0 \; , \qquad \label{BTEQN} \\
& & \hspace{-.5cm} \frac1{8 \pi} \Biggl[ -\frac{ \widetilde{B}' }{ \widetilde{r}
\widetilde{A} \widetilde{B} } \!+\! \Bigl(\frac{ \widetilde{A} \!-\! 1}{
\widetilde{r}^2 \widetilde{A} } \Bigr) \Biggr] - 
\frac{ \widetilde{\Phi}^{\prime 2} }{8\pi \alpha \widetilde{A} \widetilde{B} }
+ \frac{ \widetilde{\Phi}^2 \widetilde{F}^2}{ \widetilde{B}} + 
\frac{ \widetilde{F}^{\prime 2} }{\widetilde{A}}
- \widetilde{M}_0^2 \widetilde{F}^2 = 0 \; . \qquad \label{ATEQN}
\end{eqnarray}
(Recall that $\alpha \equiv e^2/4\pi\epsilon_0 \approx 1/137$ is the 
fine structure constant.) The kinetic and potential terms are,
\begin{eqnarray}
\widetilde{K} & = & \frac{ \widetilde{\Phi}^{\prime 2} }{8 \pi \alpha
\widetilde{A} } + \widetilde{\Phi}^2 \widetilde{F}^2 \; , \qquad \\
\widetilde{P} & = & \frac1{8\pi} \Biggl[ \frac{\widetilde{A}'}{
\widetilde{r} \widetilde{A}^{\prime 2} } \!+\! \Bigl( 
\frac{ \widetilde{A} \!-\! 1}{\widetilde{r}^2 \widetilde{A} }\Bigr) \Biggr]
- \frac{\widetilde{F}^{\prime 2}}{\widetilde{A}} - \widetilde{M}_0^2
\widetilde{F}^2 \; . \qquad
\end{eqnarray}
The normalization condition is,
\begin{equation}
2 \int_0^{\infty} \!\! d\widetilde{r} \sqrt{-\widetilde{g}} \,
\frac{ \widetilde{\Phi} \widetilde{F}^2 }{
\widetilde{B} } = 1 \; .\label{Tnorm}
\end{equation}
And the final result is,
\begin{equation}
\widetilde{M}_{\rm 0th} = \widetilde{E} - 8\pi \int_0^{\infty} \!\!
d\widetilde{r} \, \widetilde{r}^2 \sqrt{ \widetilde{A} \widetilde{K}
\widetilde{P} } \; .
\end{equation}

\begin{figure}
\includegraphics[width=6.5cm,height=6.5cm]{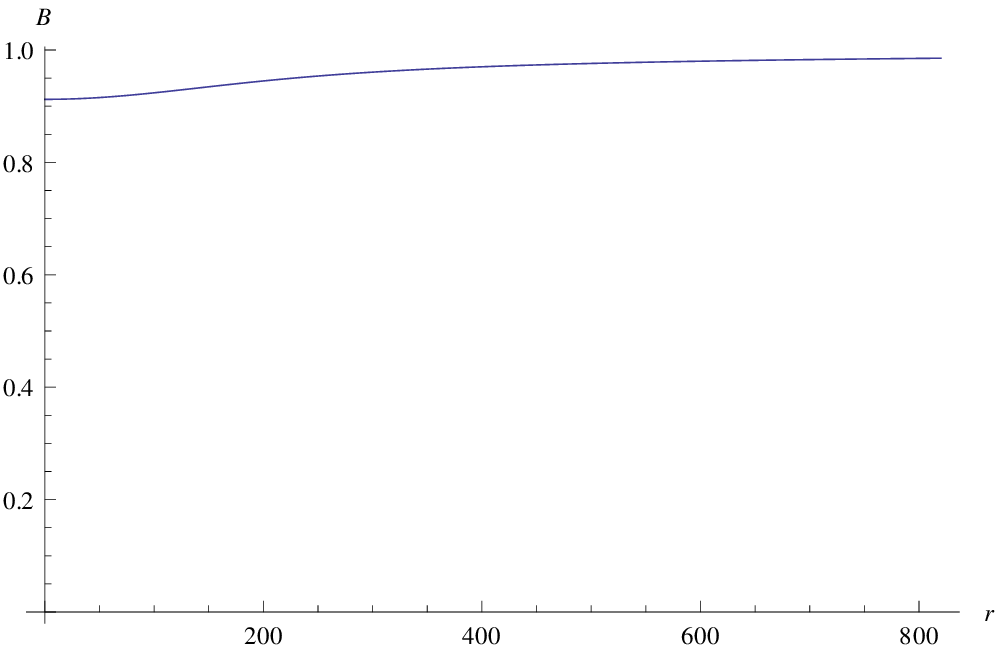}
\hspace{.5cm}
\includegraphics[width=6.5cm,height=6.5cm]{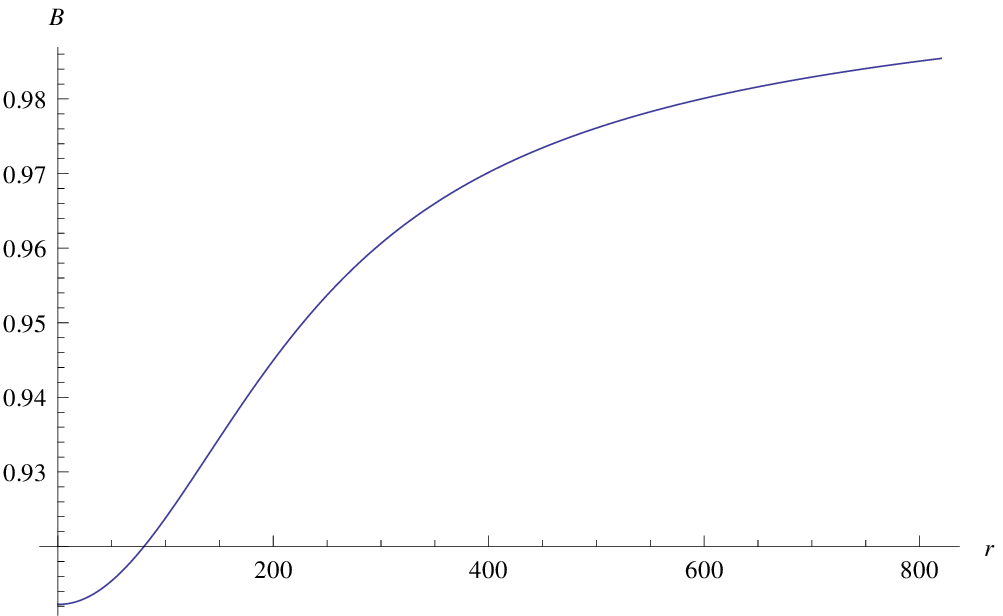}
\caption{Plots of minus the $tt$ component of the metric $B(r)$ 
(dimensionless) as a function of $r$ (in units of $1/M_{\rm Pl} = 
\sqrt{G}$). The right hand figure has an expanded vertical axis to 
show the small variation of the field. These figures were generated 
for bare mass $M_0 = 0.60 \, M_{\rm Pl}$.}
\label{Bplots}
\end{figure}

The nonlinear nature of this problem requires a special solution 
strategy. The development of our technique was facilitated by the vast
amount of work that has been done of ``boson stars'' \cite{review,JB}.
There has also been a recent study by Carlip of gravitationally bound
atoms \cite{Carlip}. 

Our strategy is to begin by evolving equations (\ref{FTEQN}-\ref{ATEQN}) 
outward from $\widetilde{r} = 0$, with arbitrary choices for 
$\widetilde{F}(0) > 0$, $\widetilde{\Phi}(0) < 0$ and $\widetilde{B}(0) 
> 0$, and with the other boundary values at,
\begin{equation}
\widetilde{F}'(0) = 0 \;\; , \;\; \widetilde{\Phi}(0) = 0 \;\; , \;\;
\widetilde{A}(0) = 1 \; .
\end{equation}
The choice of $\widetilde{B}(0) > 0$ really is arbitrary because we will
eventually make a global re-scaling of time to force
$\widetilde{B}(\widetilde{r})$ to approach one at infinity. However,
the choice of $\widetilde{\Phi}(0)$ essentially gives the energy, and
this matters of course. There is zero probability of guessing a true
eigenvalue. With the other conditions fixed, varying 
$\widetilde{\Phi}(0)$ gives solutions for which 
$\widetilde{F}(\widetilde{r})$ either becomes negative (which a
magnitude cannot do) or grows at infinity (which a normalizable
solution cannot do). One knows that a true energy eigenvalue has been 
bracketed between two different choices of $\widetilde{\Phi}(0)$ when 
the behavior of $\widetilde{F}(\widetilde{r})$ changes from one extreme 
to the other. Then one closes in on the eigenvalue to whatever
accuracy is desired. Note that this means cutting off the behavior
of the solution past a certain value of $\widetilde{r}$, beyond which
$\widetilde{F}(\widetilde{r})$ begins to degenerate.

\begin{figure}
\includegraphics[width=6.5cm,height=6.5cm]{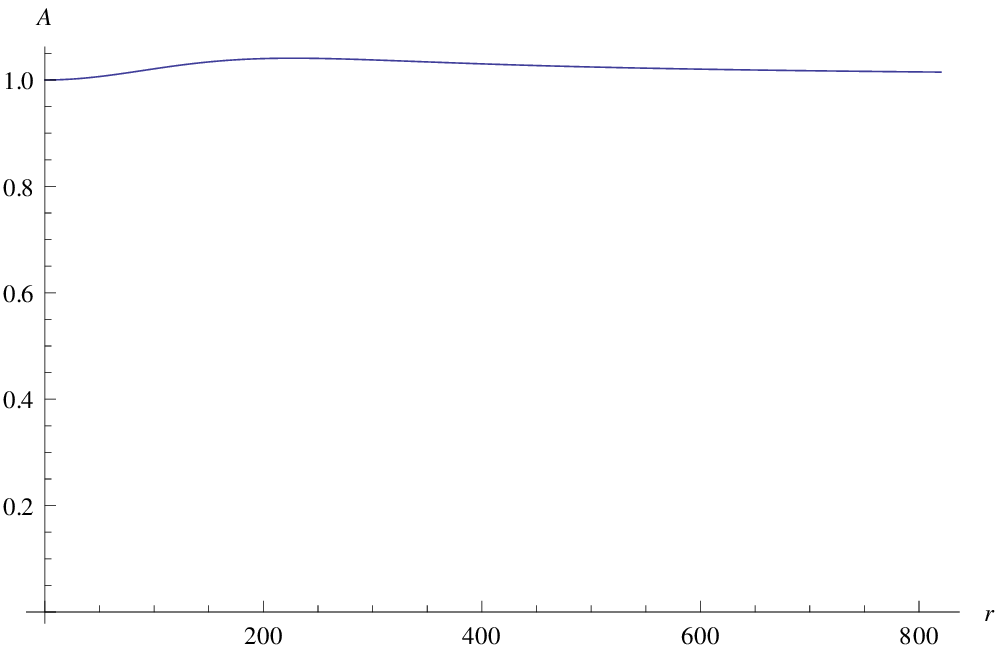}
\hspace{.5cm}
\includegraphics[width=6.5cm,height=6.5cm]{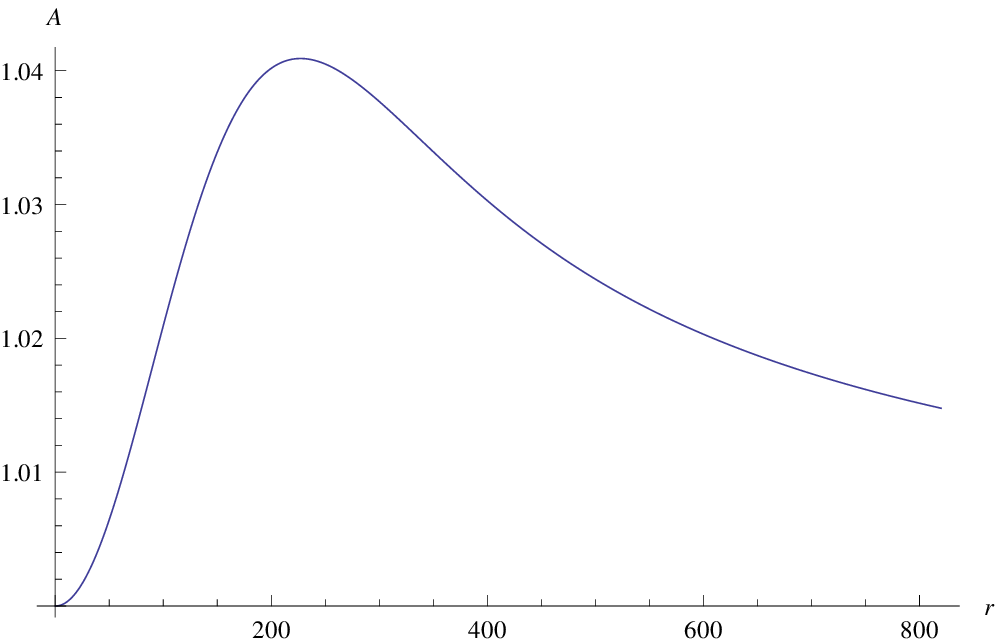}
\caption{Plots of the $rr$ component of the metric $A(r)$ (dimensionless)
as a function of $r$ (in units of $1/M_{\rm Pl} = \sqrt{G}$). The right 
hand figure has an expanded vertical axis to show the small variation 
of the field. These figures were generated for bare mass $M_0 = 0.60 \,
M_{\rm Pl}$.}
\label{Aplots}
\end{figure}

The procedure we have just outlined gives a solution which is
{\it normalizable}, but not yet normalized. For that we compute
(\ref{Tnorm}) and then either increase or decrease $\widetilde{F}(0)$
as needed. Of course the nonlinear nature of this problem means that
one does not get a solution by simply multiplying 
$\widetilde{F}(\widetilde{r})$ by a constant! We must instead start 
from the new $\widetilde{F}(0)$ and again go through the process of
trapping the energy eigenvalue. However, our evolution programs are 
efficient enough that this can be done to high accuracy, and fairly 
quickly. 

Figures~\ref{FandPhiplots}-\ref{Aplots} show the behavior of the
fields for $\widetilde{M} = 0.6$. For this bare mass the energy
is $\widetilde{E} \approx 0.586378$ and the total mass is 
$\widetilde{M} \approx -0.000882$. A measure of the numerical error
is the accuracy with which the scalar action vanishes, which is 
$\widetilde{S}_{\rm SC}/\Delta t \approx 3 \times 10^{-6}$. Another 
measure of accuracy comes from the finite cutoff at $\widetilde{r} =
\widetilde{R}_{\rm cut}$, occasioned by the finite accuracy of 
$\widetilde{E}$. For that bare mass we cut the various integrations 
off at $\widetilde{R}_{\rm cut} = 500$, which corresponds to a 
contribution of $\alpha/\widetilde{R}_{\rm cut} \approx 10^{-5}$ from 
the electromagnetic tail.

The variational results (\ref{res1}-\ref{res4}) obtained in the 
previous subsection are quite different from the numerical
solution. From Fig.~\ref{FandPhiplots} one can see than the
scalar amplitude has roughly the same shape as that of our 
trial function (\ref{Fansatz}), but with initial height
$G a \approx 0.0065$, rather than the variational value (\ref{res2})
of $G a \approx 1.7$. And the radial extent is about $\sqrt{G} \, R \approx 
300$, rather than the variational result (\ref{res1}) of $\sqrt{G} \, R
\approx 0.20$. There is no chance that the discrepancy derives from the
numerical solution, the error of which we estimate to be no larger 
than $10^{-5}$. The problem must lie instead with the variational 
formalism. Our trial solution seems roughly correct, but it may be 
that, like $B(r)$, the gravitational potential $A(r)$ represents a 
negative energy direction in field space. In that case minimizing the
constrained action would take us away from the actual solution,
which seems to be what has happened.

\begin{table}

\vbox{\tabskip=0pt \offinterlineskip
\def\tablerule{\noalign{\hrule}}
\halign to390pt {\strut#& \vrule#\tabskip=1em plus2em&
\hfil#\hfil& \vrule#& \hfil#\hfil& \vrule#&
\hfil#\hfil& \vrule#& \hfil#\hfil& \vrule#&
\hfil#\hfil& \vrule#\tabskip=0pt\cr
\tablerule
\omit&height4pt&\omit&&\omit&&\omit&&\omit&&\omit&\cr
\omit&height2pt&\omit&&\omit&&\omit&&\omit&&\omit&\cr
&& $\!\!\!\! \widetilde{M}_0 \!\!\!\!$
&& $\!\!\!\! \widetilde{E} \!\!\!\!$
&& $\!\!\!\! \widetilde{M} \!\!\!\!$
&& $\!\!\!\! \widetilde{S}_{\rm SC}/\Delta t \!\!\!\!$
&& $\!\!\!\! \widetilde{R}_{\rm cut} \!\!\!\!$ & \cr
\omit&height4pt&\omit&&\omit&&\omit&&\omit&&\omit&\cr
\tablerule
\omit&height2pt&\omit&&\omit&&\omit&&\omit&&\omit&\cr
&& 0.15 && 0.149994 && -0.000320 && $2.0 \times 10^{-10}$ && 70,000 & \cr
\omit&height2pt&\omit&&\omit&&\omit&&\omit&&\omit&\cr
\tablerule
\tablerule
\omit&height2pt&\omit&&\omit&&\omit&&\omit&&\omit&\cr
&& 0.20 && 0.199962 && -0.000260 && $1.5 \times 10^{-9}$ && 25,000 & \cr
\omit&height2pt&\omit&&\omit&&\omit&&\omit&&\omit&\cr
\tablerule
\omit&height2pt&\omit&&\omit&&\omit&&\omit&&\omit&\cr
&& 0.25 && 0.249870 && -0.000505 && $2.9 \times 10^{-9}$ && 11,750 & \cr
\omit&height2pt&\omit&&\omit&&\omit&&\omit&&\omit&\cr
\tablerule
\omit&height2pt&\omit&&\omit&&\omit&&\omit&&\omit&\cr
&& 0.30 && 0.299653 && -0.000239 && $9.6 \times 10^{-9}$ && 4,500 & \cr
\omit&height2pt&\omit&&\omit&&\omit&&\omit&&\omit&\cr
\tablerule
\omit&height2pt&\omit&&\omit&&\omit&&\omit&&\omit&\cr
&& 0.35 && 0.349215 && -0.001075 && $5.7 \times 10^{-8}$ && 3,200 & \cr
\omit&height2pt&\omit&&\omit&&\omit&&\omit&&\omit&\cr
\tablerule
\omit&height2pt&\omit&&\omit&&\omit&&\omit&&\omit&\cr
&& 0.40 && 0.398424 && -0.000455 && $1.3 \times 10^{-7}$ && 1,800 & \cr
\omit&height2pt&\omit&&\omit&&\omit&&\omit&&\omit&\cr
\tablerule
\omit&height2pt&\omit&&\omit&&\omit&&\omit&&\omit&\cr
&& 0.45 && 0.447073 && -0.000826 && $3.5 \times 10^{-7}$ && 1,300 & \cr
\omit&height2pt&\omit&&\omit&&\omit&&\omit&&\omit&\cr
\tablerule
\omit&height2pt&\omit&&\omit&&\omit&&\omit&&\omit&\cr
&& 0.50 && 0.494904 && -0.000788 && $7.3 \times 10^{-7}$ && 1,400 & \cr
\omit&height2pt&\omit&&\omit&&\omit&&\omit&&\omit&\cr
\tablerule
\omit&height2pt&\omit&&\omit&&\omit&&\omit&&\omit&\cr
&& 0.55 && 0.541546 && -0.000412 && $1.6 \times 10^{-6}$ && 1,100 & \cr
\omit&height2pt&\omit&&\omit&&\omit&&\omit&&\omit&\cr
\tablerule
\omit&height2pt&\omit&&\omit&&\omit&&\omit&&\omit&\cr
&& 0.60 && 0.586378 && -0.000815 && $3.4 \times 10^{-6}$ && 500 & \cr
\omit&height2pt&\omit&&\omit&&\omit&&\omit&&\omit&\cr
\tablerule
\omit&height2pt&\omit&&\omit&&\omit&&\omit&&\omit&\cr
&& 0.65 && 0.628511 && -0.001440 && $8.1 \times 10^{-6}$ && 350 & \cr
\omit&height2pt&\omit&&\omit&&\omit&&\omit&&\omit&\cr
\tablerule
\omit&height2pt&\omit&&\omit&&\omit&&\omit&&\omit&\cr
&& 0.70 && 0.666426 && -0.001866 && $1.7 \times 10^{-5}$ && 275 & \cr
\omit&height2pt&\omit&&\omit&&\omit&&\omit&&\omit&\cr
\tablerule
\omit&height2pt&\omit&&\omit&&\omit&&\omit&&\omit&\cr
&& 0.75 && 0.696992 && -0.000532 && $4.1 \times 10^{-5}$ && 400 & \cr
\omit&height2pt&\omit&&\omit&&\omit&&\omit&&\omit&\cr
\tablerule}}

\caption{Numerical results for the scalar energy $\widetilde{E}$ and
the total mass $\widetilde{M}$ for different values of the bare mass
$\widetilde{M}_0$. Also given are the scalar action $\widetilde{S}_{
\rm SC}/\Delta t$, which should vanish, and the cutoff radius beyond 
which the finite accuracy of the energy eigenvalue makes the solution 
unreliable. All quantities are expressed in Planck units.}

\label{results}

\end{table}

Table~\ref{results} gives our results for $\widetilde{E}$ and
$\widetilde{M}$ for a variety of different bare masses. The most
obvious feature is the almost total cancellation between the
energy of the scalar wave function and the field action, to give
a very small, negative total mass. This is physical nonsense 
because it fails to agree with the mass one can read off from
asymptotic values of the metric. We believe that the problem
arises from the asymptotic conditions (\ref{asymp1}-\ref{asymp2})
--- which are certainly valid for scattering with other particles
--- not being right for the study of self-interactions. We believe
that this can be fixed without much change.

The other features of our numerical work are:
\begin{itemize}
\item{The energy $\widetilde{E}$ agrees with the mass inferred from
the asymptotic values of the metric.}
\item{There is no bound state unless the bare mass $\widetilde{M}_0$ 
exceeds the ADM result of $\sqrt{\alpha} \approx 0.85$ \cite{JB}.}
\item{The bound state energy $\widetilde{E}$ is in all cases less 
than the bare mass.}
\item{The ratio $\widetilde{E}/\widetilde{M}_0$ increases with 
$\widetilde{M}_0$ and eventually becomes zero \cite{JB}.}
\end{itemize}
Because there would not even be any bound states without gravity,
it seems fair to conclude that the system depends nonanalytically
upon $G$.

\section{Epilogue}

We have explored the possibility that the apparent problems of
quantum general relativity may be artifacts of conventional
perturbation theory. One might think this unlikely because the
absence of recognizable, low energy quantum gravitational phenomena
implies that {\it some} asymptotic series expansion is wonderfully
accurate. However, it may be that the correct series 
involves logarithms or fractional powers of Newton's constant. If
that were the case, trying to re-expand in integer powers of $G$
would result in an escalating series of divergences, which is
exactly what conventional perturbation theory shows.

We studied this possibility in the context of computing the mass of
a charged, gravitating scalar. An exact result for the classical
limit of this system was derived by ADM in 1960 \cite{ADM1}, and it
does exhibit both nonanalytic dependence upon $G$ and the breakdown
of conventional perturbation theory. If the classical point particle
is regulated to be a spherical shell of radius $R$, the ADM result
is,
\begin{equation}
M_R = \frac{R c^2}{G} \sqrt{1 \!+\! \frac{2 G M_0}{R c^2} \!+\!
\frac{e^2 G}{4\pi \epsilon_0 R^2 c^4} } \, - \frac{R c^2}{G} \; .
\end{equation}
The correct zero radius limit is $M = \sqrt{\alpha/G}$.
Its finiteness results from negative gravitational interaction
energy canceling the positive electromagnetic energy. In contrast,
the perturbative result is obtained by first expanding the square
root in powers of $G$ and $e^2$, which produces a series of
ever-higher divergences with alternating signs. The alternating
signs are a signal that gravity is trying to cancel the
electromagnetic self-energy divergence, but this cancellation can
never happen in conventional perturbation theory because the
gravitational response to a divergence at one order is delayed until
one order higher. What we need for quantum gravity is an
alternate expansion in which the negative gravitational interaction
energy has a chance to ``keep up'' with what is going on in the
positive energy sectors.

In section 3 we derived an exact functional integral expression
(\ref{finalform}) for the scalar mass. We then developed an alternate
asymptotic expansion based on the Method of Stationary Phase, with
the full functional integrand --- not just the action --- used to
determine the stationary point. This is more difficult to implement
than conventional perturbation theory, but it is also more correct. A
simple integral representation for the Bessel function illustrates
the distinction between our approach and that of conventional
perturbation theory,
\begin{equation}
J_N(z) = \frac1{2\pi} \int_{-\pi}^{\pi} \!\!\! d\theta \, e^{i z
\sin(\theta)} \times \Bigl(e^{-i \theta} \Bigr)^N \; .
\end{equation}
In our approach both factors are included in the exponent and the
two stationary points are found by minimizing the function $f(\theta) =
z \sin(\theta) - N \theta$,
\begin{equation}
f'(\theta_{\pm}) = 0 \qquad \Longrightarrow \qquad \theta_{\pm} = \pm
{\rm acos}\Bigl( \frac{N}{z}\Bigr) \; .
\end{equation}
The values of the function and its second derivative at these
points are,
\begin{equation}
f(\theta_{\pm}) = \pm \Bigl[ \sqrt{z^2 \!-\! N^2} - N {\rm acos}\Bigl(
\frac{N}{z}\Bigr) \Bigr] \qquad , \qquad f''(\theta_{\pm}) =
\mp \sqrt{z^2 \!-\! N^2} \; .
\end{equation}
And the result for the 0th and 1st order contributions is,
\begin{equation}
J_N(z) \longrightarrow \sqrt{\frac{2}{\pi \sqrt{z^2 \!-\! N^2}} }
\cos\Bigl[ \sqrt{z^2 \!-\! N^2} - N {\rm acos}\Bigl( \frac{N}{z}\Bigr)
\!-\! \frac{\pi}4 \Bigr] \; .
\end{equation}
In contrast, conventional perturbation theory would be based on
the function $f(\theta) = z \sin(\theta)$, with the stationary points
at $\theta_{\pm} = \pm \frac{\pi}2$. The result for the 0th and 1st
order contributions from conventional perturbation theory is,
\begin{equation}
J_N(z) \longrightarrow \sqrt{\frac{2}{\pi z}} \cos\Bigl[z \!-\!
N \frac{\pi}2 \!-\! \frac{\pi}4 \Bigr] \; .
\end{equation}

Section 4 presents an analysis of the new expansion in the context of
a simplified model. We conclude that all the old $\ell$ loop diagrams
appear at $\ell$-th order in the new expansion. However, the old
$\ell$ loop diagrams are combined with an infinite class of new
diagrams which possess more external lines and no more than $\ell$
loops. The new diagrams which are added at $\ell$-th order are all
subtracted at higher orders, so we are really adding zero to the usual
expansion. Because the new $\ell$-th order diagrams have no more than
$\ell$ loops, the divergences of the new expansion can be no worse
than those of conventional perturbation theory. Because infinitely 
many new diagrams are added at each order, the new expansion can
depend nonanalytically on Newton's constant. It also offers a way
in which the negative gravitational interaction energy can respond,
at the same order, to problems in the positive energy sectors. These 
are all desirable features, although it must be admitted that these 
is no guarantee at this stage that the new expansion is any better 
than the old one.

The analysis of section 4 was done only to understand how the new
expansion compares with the old one. There are much better ways of
actually implementing the new expansion. We exploit two of these
methods in section 5 to evaluate the zeroth order result. Our
analysis is based on interpreting the zeroth order term as the
phase developed by a first-quantized Klein-Gordon scalar moving in
the gravitational and electrodynamic potentials which are sourced
by its own probability current. The fact that this system reduces
to the ADM problem for $\hbar \rightarrow 0$ provides a solid reason
for believing both that the negative energy gravitational interactions
cancel at least some of the usual self-energy divergences, and that
the final result depends nonanalytically on Newton's constant.

Evaluating the zeroth order term of the new expansion amounts to
solving for a bound state of the scalar in its own potentials.
Although we cannot obtain exact solutions for all four of the relevant 
field equations, we were able to eliminate one of the negative energy 
gravitational degrees of freedom to derive a variational formalism. 
We were also able to solve the equations numerically, taking 
advantage of the vast body of work which has been done on
``boson stars'' \cite{review,JB}.

We achieved high numerical accuracy which revealed a substantial
discrepancy with the variational approach. This probably means that
the gravitational potential we were not able to eliminate also
carries negative energy, so that minimizing the constrained action
takes one away from the true solution. As was seen in previous
numerical work, we found that there are no solutions unless the
bare mass $M_0$ is greater than the ADM result of $\sqrt{\alpha/G}$.
We developed solutions for many choices of $M_0$ above this limit.
All of them show an almost total cancellation between the energy of 
the scalar wave function and the field energy of the gravitational 
and electromagnetic potentials. This gives nearly zero for the total 
mass, which seems to be nonsense. It also fails to agree with a 
determination of the scalar mass from the asymptotic values of the 
gravitational potentials.

The problem seems to derive from our use of the asymptotic conditions 
(\ref{asymp1}-\ref{asymp2}). Expressed in simple words, these conditions 
mean that ``the fields become free at asymptotically early and late 
times''. That is perfectly true (in the weak operator sense and 
assuming the existence of a mass gap) for interactions {\it between} 
different particles, which is the usual application \cite{LSZ}. 
However, we are here trying to use the conditions to study interactions 
of a particle {\it with itself}. These self-interactions would usually 
be subsumed into forcing the field strength and mass to come out right 
by renormalization, but that is exactly what we are not doing. We
believe that when a more accurate procedure is used to interpolate the
single particle states --- which might be as simple as including a $U(1)$
gauge string between the two fields to make them invariant --- then 
the nonsense result for $M$ will go away, and most of our analysis of
the new expansion will be unchanged.

\centerline{\bf Acknowledgements}

We are grateful to Stanley Deser for years of guidance and
inspiration. We have profited from conversations on this subject
with G. T. Horowitz and T. N. Tomaras. This work was partially
supported by European Union Grant
FP-7-REGPOT-2008-1-CreteHEPCosmo-228644, by FQXi Grant RFP2-08-31,
by NSF grant PHY-0855021, and by the Institute for Fundamental
Theory at the University of Florida.

\end{document}